\documentclass[12pt,a4paper]{article}
\usepackage{ytableau}
\usepackage{jheppub}
\usepackage{multirow}
\usepackage{chemarrow}
\usepackage{amsmath,amssymb,amsfonts,mathtools}
\usepackage{float}
\usepackage{subfig}
\usepackage{scalefnt}
\usepackage{wrapfig}
\usepackage{fancybox}
\usepackage{ifsym}
\usepackage{relsize}
\newcommand{\bea}{\begin{eqnarray}\displaystyle}
\newcommand{\eea}{\end{eqnarray}}
\newcommand{\nn}{\nonumber}
\newlength{\arrow}
\newcommand\half{\frac{1}{2}}

{\setlength{\fboxsep}{15pt}
\setlength{\mylength}{\linewidth}%
\addtolength{\mylength}{-2\fboxsep}%
\addtolength{\mylength}{-2\fboxrule}%
\Sbox
\minipage{\mylength}%
\setlength{\abovedisplayskip}{0pt}%
\setlength{\belowdisplayskip}{0pt}%
\equation}%
{\endequation\endminipage\endSbox
\[\fbox{\TheSbox}\]}

\begin{document}
\title{E + E $\rightarrow$ H}
\author[\ast\ddagger]{Babak Haghighat,}
\author[\ast]{Guglielmo Lockhart,}
\author[\ast]{Cumrun Vafa}
\affiliation[\ast]{Jefferson Physical Laboratory, Harvard University, Cambridge, MA 02138, USA}
\affiliation[\ddagger]{Department of Mathematics, Harvard University, Cambridge, MA 02138, USA}

\abstract{E-strings arise from M2 branes suspended between an M5 brane and an M9 plane.  In this paper
we obtain explicit expressions for the elliptic genus of two E-strings using a series of string dualities.  Moreover we show how this can be used to recover the elliptic genus of two $E_8\times E_8$ heterotic strings using the Ho\v rava-Witten realization of heterotic strings in M-theory.  This involves highly non-trivial identities among Jacobi forms, and is remarkable in light of the fact that E-strings are `sticky' and form bound states whereas heterotic strings do not form bound states.}
\maketitle

\ytableausetup{mathmode, boxsize=0.4em}

\section{Introduction}
\label{sec:1}

The 6d SCFTs with (1,0) supersymmetry are among the least understood quantum field theories.  This
is partly due to the fact that they have light (tensionless) strings as a main ingredient.  In previous work \cite{Haghighat:2013gba,Haghighat:2013tka} (see also \cite{Hohenegger:2013ala,Schwarz:2013bua})
we have made some progress in understanding the relation between the supersymmetric partition function of these theories
and the partition function of the associated strings.  In particular as was shown in \cite{Iqbal:2012xm,Kallen:2012cs,Lockhart:2012vp,Imamura:2012bm,Kim:2012qf,Nieri:2013yra} the partition function of these theories on $S^4\times S^1$ or $S^5$ can be computed by the partition function of these theories on $\mathbb{R}^4 \ltimes T^2$
backgrounds (where $\mathbb{R}^4$ is twisted as we go along the $T^2$ cycles and Wilson lines are turned on for the various global symmetries). This in turn can be computed by the elliptic genus of the tensionless strings on $T^2$, as they constitute the only BPS instantons of these theories.  In particular, for a theory that has $k$ different types of strings with tensions $(t_1, \dots, t_k)$ (which
can be identified with scalar vevs in the associated tensor multiplets), one has
$$Z^{(1,0)}_{\mathbb{R}^4\ltimes T^2}=\sum_{\vec{n}} e^{-\vec{n}\cdot \vec{t}}\, Z^{\vec{n}}_{T^2}~ ,$$
where $Z^{\vec{n}}_{T^2} $ denotes the elliptic genus of a collection of $(n_1, \dots, n_k)$ strings.  Furthermore, this
partition function can be identified with either the refined topological string partition function of a dual geometry, or if there is a gauge theory
description, with the Nekrasov partition function, and one can in principle use methods developed in those contexts to study it.  This can be used to compute the elliptic genus of tensionless strings.  On the other hand, in some cases
(such as the (1,0) theory obtained by probing the $A_{N}$ singularity with M5 branes
 \cite{Haghighat:2013tka}) it is possible to reverse this by identifying the theory on the tensionless strings (which is in some cases given by a quiver gauge theory) and using
it to compute the partition function of the (1,0) SCFT itself.\\

\noindent The main focus of this paper is on E-strings, which arise \cite{Witten:1995gx,Ganor:1996mu,Seiberg:1996vs} from an M5 brane probing the Ho\v rava-Witten M9 plane; the E-strings are identified with M2 branes stretched between the M5 brane and the M9 plane. In this context string dualities \cite{Klemm:1996hh} relate this system to topological strings on a CY 3-fold in the vicinity of the ${1\over 2}K3$ surface. The topological string partition function for this theory has been studied in \cite{Klemm:1996hh,Minahan:1998vr,Hosono:1999qc,Mohri:2001zz,Eguchi:2002fc,Iqbal:2002rk,Eguchi:2002nx,Sakai:2011xg,Huang:2013yta} and, even though major progress has been made, the result is still incomplete.  Our main aim is to build on these partial results to obtain explicit  formulas for the elliptic genus of two E-strings, $Z^{\textrm{E-str}}_{2}$; the answer we propose passes highly non-trivial checks\footnote{The result for 1 E-string is much simpler and was already studied in \cite{Klemm:1996hh,Hosono:1999qc}.}. We find that, as in the case
of M-strings \cite{Haghighat:2013gba}, two E-strings have a rather non-trivial bound state structure, unlike fundamental strings which do not form bound states.  The lack of the bound states for fundamental strings is reflected in the fact that the partition function
for $n$ fundamental strings is simply an order $n$ Hecke transform of the one for a single string.  We find that this is not the case for two E-strings.\\

\noindent This raises the following question:  We know that an M5 brane placed between the two M9 planes of M-theory gives rise to $E_L$-strings 
from the left plane as well as $E_R$-strings from the right plane.  On the other hand, we also know that an $E_8\times E_8$ heterotic string can be identified
with an M2 brane stretched between the two M9 planes \cite{Horava:1996ma}.   Thus, $n$ pairs of E-strings can recombine to give $n$ heterotic strings ($H$):
$$nE_L+nE_R\rightarrow nH~.$$
\begin{figure}[here!]
  \centering
	\includegraphics[width=0.8\textwidth]{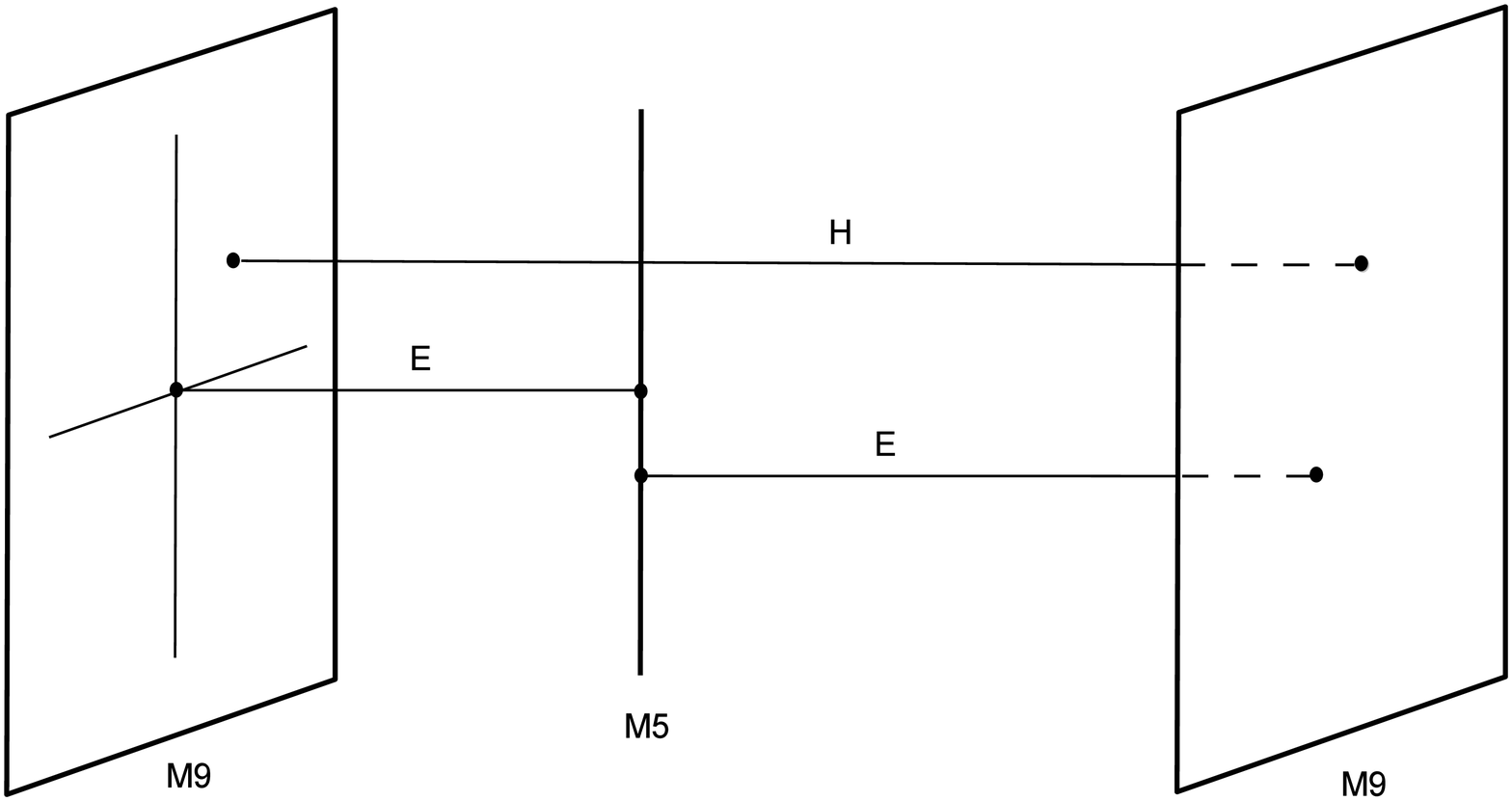}
\end{figure}
At first glance this is puzzling, as it is not obvious how the lack of bound states of two heterotic strings is compatible with the existence of a bound state structure for two E-strings.
The answer to this puzzle is provided by the presence of the M5 brane which serves as a `glue' for the M2 branes. One may also wonder whether it is possible to recover the partition function of heterotic strings from that of E-strings.  The fact that E-strings recombine to give heterotic strings strongly suggests that this should be possible, and in this paper we indeed show that this can be done at least up to $n=2$ E-strings.
The basic idea is to view the theory of $n$ M2 branes on $\mathbb{R}\times T^2$, in the limit where the area of the $T^2$ (on which the elliptic genus does not depend)  is small, as a quantum mechanical system on $\mathbb{R}$.   Under this reduction the states in the Hilbert space of $n$ M2 branes are labelled by Young diagrams of size $n$ \cite{Haghighat:2013gba,Kim:2010mr}, and M5 branes as well as M9 planes intersecting the M2 branes on $T^2$ can be interpreted as operators or states in this quantum mechanical system. We call them domain wall operators/states due to their interpretation in the worldvolume theory of M2 branes. In a previous paper \cite{Haghighat:2013gba} we computed the contribution of M5 brane domain walls to this quantum mechanical system.  Here, using low genus results from topological strings for up to two E-strings, and using the known M5 brane domain wall, we determine the exact M9 domain wall wave function for up to two M2 branes. We then deduce a closed formula for the elliptic genus of two E-strings, which from the viewpoint of topological string theory provides an all-genus A-model amplitude for up to two E-strings.  We also test our M9 domain wall expressions by checking whether the left and right walls combine correctly into the elliptic genus of up to 2 heterotic strings, and remarkably we find that they do (up to taking into account a symmetrization which the heterotic string enjoys, and which is broken in the E-string background by the M5 brane).\newline

\noindent The organization of this paper is as follows: In Section \ref{sec:2} we present the M2-M5-M9 configurations corresponding to the heterotic, E- and M-strings. In Section \ref{sec:3} we review the computation of the M-string elliptic genus in terms of M5 domain wall operators and the resulting partition function for two M5 branes. In Section \ref{sec:4} we obtain the elliptic genus of heterotic strings by using the Hecke transform. We then proceed in Section \ref{sec:5} to outline the series of string dualities which relate the E-string theory to the topological string on the half-K3 Calabi-Yau threefold. Finally, in Section \ref{sec:6} we determine the M9 domain wall operator for up to two strings and use it to compute the elliptic genus of E- and heterotic strings.

\section{M2 branes on $T^2 \times \mathbb{R}$ and boundary conditions}
\label{sec:2}

In this section we review possible boundary conditions for M2 branes together with the preserved supersymmetries. To do this we consider M-theory on $T^2 \times \mathbb{R}^9$ and take the M2 branes to wrap the $T^2$ and extend along one of the directions of $\mathbb{R}^9$, so that their worldvolume is given by $T^2 \times \mathbb{R}$. We choose coordinates $X^I,~I=0,1,2,\cdots,10$ and parametrize the torus by $X^0,~X^1$ and take the direction along which the M2 branes are extended to be $X^6$. We obtain different boundary conditions by letting the M2 branes end on M5 branes or M9 planes. This can be done in various combinations which we describe here. 

\subsection*{M9-M9}

\begin{figure}[here!]
  \centering
	\includegraphics[width=1.0\textwidth]{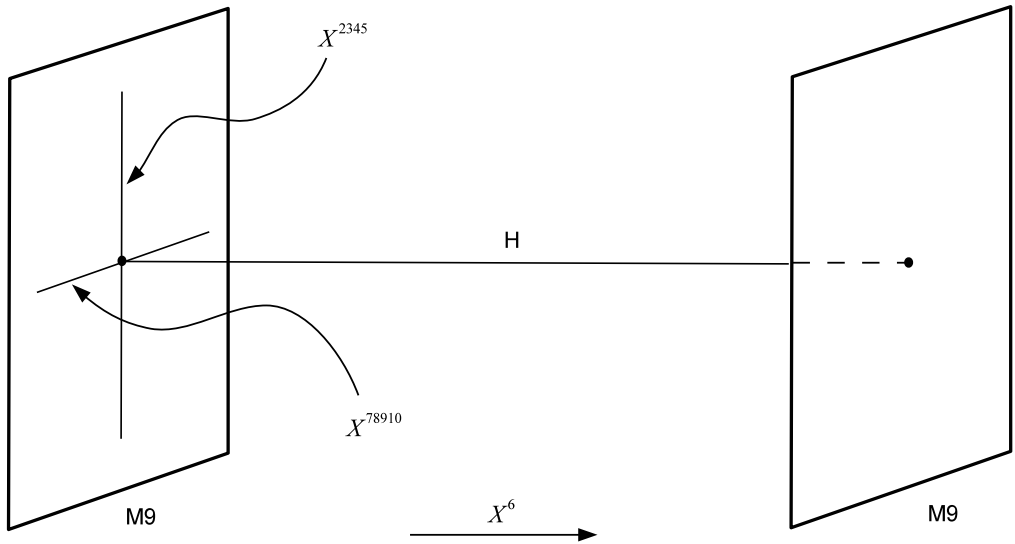}
  \caption{An M2 branes suspended between M9 planes corresponding to the heterotic string. The worldvolume of the M2 branes and M9 planes share a common $T^2$ which is suppressed in the picture. The directions orthogonal to the torus are represented as the separation $X^6$ and the quaternionic subspaces $X^{2345}$ and $X^{78910}$.}
  \label{fig:M9-M9}
\end{figure}

Here the relevant setup is the one of Ho\v rava and Witten \cite{Horava:1996ma}. We compactify M-theory on $T^2 \times \mathbb{R}^8 \times S^1/\mathbb{Z}_2$ where the $\mathbb{Z}_2$ acts as an orbifold action,
\begin{equation}
X^6 \mapsto -X^6,
\end{equation}
together with a suitable action on the fields. At the two fixed points of the orbifold action, $X^6=0$ and $X^6=\pi$, one has two fixed planes which we denote as M9 planes and are here of the topology $T^2 \times \mathbb{R}^8$; the situation is illustrated in Figure \ref{fig:M9-M9}.

\noindent In the limit where the size of $ S_1/\mathbb{Z}_2 $ goes to zero, the M2 branes give rise to heterotic strings charged under an $E_8 \times E_8$ current algebra, with each $E_8$ coming from one M9 plane \cite{Horava:1996ma}. Next, we want to look at the preserved supersymmetries on these strings. Each brane type projects out half of the $32$ supercharges as follows,
\begin{equation}
	M9:~\Gamma^6 \epsilon = \epsilon, \quad M2: ~\Gamma^{016} \epsilon = \epsilon,
\end{equation}
and thus we see that the wordsheet theory on the strings is chiral and carries $(8,0)$ supersymmetry. We can break this supersymmetry down to $(4,0)$ and $(2,0)$ by introducing a twisted background, \textit{i.e.}\!\! turning on fugacities when going along the cycles of the $T^2$. The way this works is as follows. As explained in \cite{Haghighat:2013gba} viewing the torus as $S^1 \times S^1$ we twist the $\mathbb{R}^4_{2345} \times \mathbb{R}^4_{78910}$ by the action of the Cartan subalgebra of the $SO(8)$ R-symmetry parametrized by $U(1)_{\epsilon_1}\times U(1)_{\epsilon_2}\times U(1)_{\epsilon_3} \times U(1)_{\epsilon_4}$ as we go around the cycles of the torus:
\begin{eqnarray}
	\prod_{i=1}^4 U(1)_{\epsilon_i} & : & (z_1,z_2) \mapsto (e^{2\pi i\epsilon_1} z_1,e^{2\pi i \epsilon_2}z_2), \\
	~ & : & (w_1,w_2) \mapsto (e^{2\pi i \epsilon_3}w_1,e^{2\pi i \epsilon_4}w_2),
\end{eqnarray}
where we impose the following relation
\begin{equation}
	\epsilon_1 + \epsilon_2 + \epsilon_3 + \epsilon_4 = 0,
\end{equation}
in order to preserve supersymmetry. For generic values of the $\epsilon_i$ only a $(2,0)$ subset of the supercharges is preserved which enhances to $(4,0)$ for the locus given by $\epsilon_2 = - \epsilon_1$ and $\epsilon_3 = - \epsilon_4$ or permutations of these. \\

\noindent In this paper we will be interested in the computation of the elliptic genus of $n$ heterotic strings wrapping the $T^2$, which is given by
\begin{equation}
	\textrm{Tr}_{\textrm{R}} (-1)^F \bar{q}^{H_L} q^{H_R} \prod_a x_a^{K_a},
\end{equation}
where the $K_a$ denote the Cartan generators associated with general supersymmetry preserving $SO_R(8)$ spacetime twists and $E_8 \times E_8$ fugacities. We will denote this quantity by
\begin{equation}
	Z^{\textrm{Het}}_n(\tau, \epsilon_1, \epsilon_2, \epsilon_3, \epsilon_4, \vec{m}_{E_8\times E_8}),
\end{equation}
where $\tau$ denotes the complex structure of the torus.

\subsection*{M9-M5}

\begin{figure}[here!]
  \centering
	\includegraphics[width=1.0\textwidth]{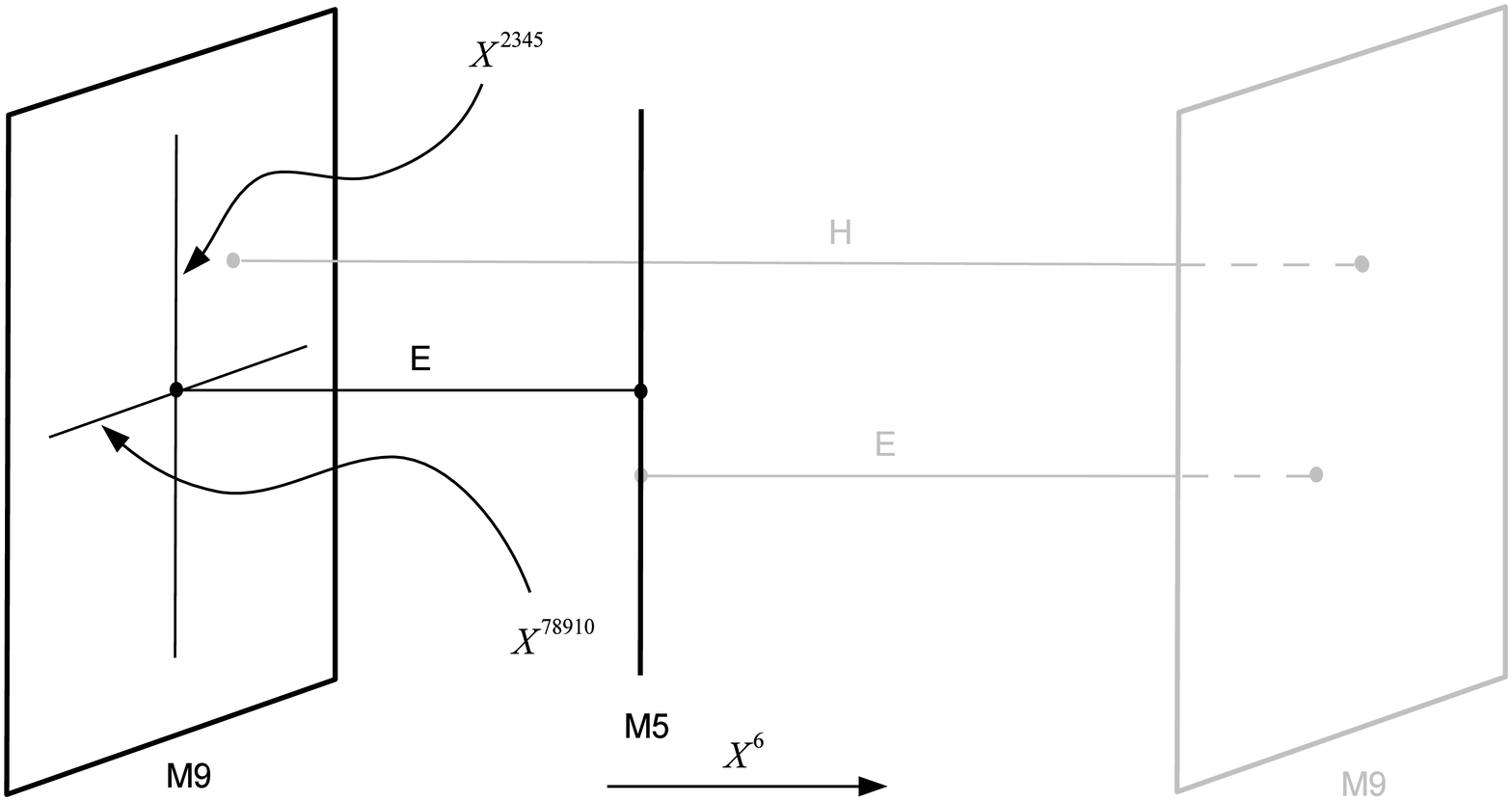}
  \caption{An M2 brane suspended between an M9 and an M5 brane corresponding to the E-string. The worldvolume of the branes share a common $T^2$ which is suppressed in the picture. The directions orthogonal to the torus are represented as the separation $X^6$ and the quaternionic subspaces $X^{2345}$ and $X^{78910}$.}
  \label{fig:M9-M5}
\end{figure}

This setup leads to the theory of E-strings \cite{Seiberg:1996vs, Ganor:1996mu} which is a six-dimensional superconformal field theory with $(1,0)$ supersymmetry. This theory arises from a system of M9 and M5 branes with M2 branes suspended between them  \cite{Ganor:1996mu}. To be more specific, we take an M9 plane as before along the coordinates $X^0,\cdots,X^5,X^7,\cdots, X^{10}$ and an M5 brane along the directions $X^0,\cdots,X^5$ and separate them along the $X^6$ direction. We depict this in Figure \ref{fig:M9-M5}.\\

\noindent Each of the branes projects out half of the $32$ supercharges and the surviving supercharges satisfy the condition
\begin{equation}
	M9:~\Gamma^6 \epsilon = \epsilon, \quad M5:~\Gamma^{012345} \epsilon = \epsilon, \quad M2:~\Gamma^{016}\epsilon = \epsilon.
\end{equation}
Thus the worldvolume theory of the E-string being the intersection of the M2 brane and the M5 brane has $(4,0)$ supersymmetry. As the M2 brane is ending only on one of the M9 planes the string is charged under one $E_8$ current algebra. One can now again consider a twisted background by introducing boundary conditions labelled by Cartan generators of $SO_R(8)$ along cycles of the $T^2$. Generic twists will break the supersymmetry down to $(2,0)$, while setting $\epsilon_1 = - \epsilon_2$ gives an enhancement to $(4,0)$. In this paper we will be interested in the computation and properties of the elliptic genus of $n$ E-strings with various fugacities turned on, namely
\begin{equation}
	Z_n^{\textrm{E-str}}(\tau, \epsilon_1,\epsilon_2,\vec{m}_{E_8}).
\end{equation}
Here it is important to note that the E-string elliptic genus does not depend on $\epsilon_3$ and $\epsilon_4$. The reason is that the six-dimensional E-string theory only enjoys a $SU(2)$ R-symmetry which can be identified with $SU(2)_L$ in the decomposition
\begin{equation}
	Spin(4)_{78910} = SU(2)_L \times SU(2)_R,
\end{equation}
while the $U(1)$ symmetry associated to $\epsilon_3-\epsilon_4$ lies in $SU(2)_R$.

\subsection*{M5-M5}

\begin{figure}[here!]
  \centering
	\includegraphics[width=1.0\textwidth]{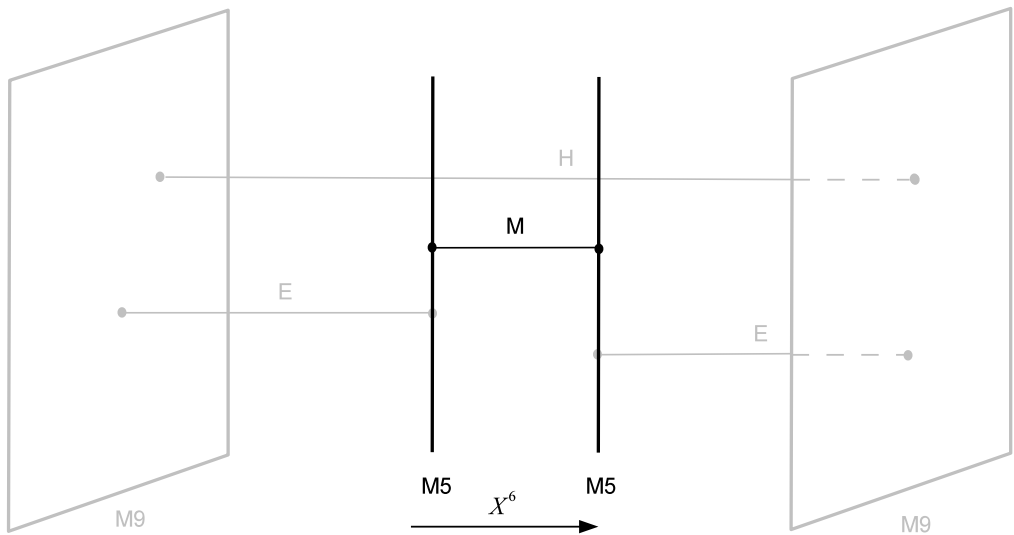}
  \caption{An M2 brane suspended between two M5 branes corresponding to the M-string. The worldvolume of the branes share a common $T^2$ which is suppressed in the picture. The M5 branes are extended along the $X^{012345}$ directions.}
  \label{fig:M5-M5}
\end{figure}

This configuration leads to the six-dimensional $A_{N-1} $ $(2,0)$ superconformal field theory \cite{Witten:1995zh}. Specializing to two M5 branes we obtain the $A_1$ theory that we describe here. The compactification of this theory on $T^2$ gives rise to $\mathcal{N}=4$ SYM in four dimensions. Taking the M5 branes to be extended along $X^{012345}$ and the M2 branes along $X^{016}$ we obtain the schematic picture shown in Figure \ref{fig:M5-M5}.

\noindent We denote by M-strings the strings arising from M2 branes stretching between the M5 branes. The supercharges preserved by M-strings obey the constraints
\begin{equation}
	M2:~ \Gamma^{016} \epsilon = \epsilon, \quad M5:~\Gamma^{012345} \epsilon = \epsilon,
\end{equation}
which lead to $(4,4)$ supersymmetry on the string worldsheet. Turning on fugacities $\epsilon_i$ breaks this down to $(4,0)$, $(2,2)$, or $(2,0)$ SUSY as discussed in \cite{Haghighat:2013gba}. The theory one arrives at is five-dimensional $\mathcal{N}=2^*$ SYM compactified on a circle. The mass $m$ of the adjoint hypermultiplet is related to the $\epsilon_3$ and $\epsilon_4$ parameters as follows:
\begin{equation}
	\epsilon_3 = -m - \frac{\epsilon_1 + \epsilon_2}{2}, \quad \epsilon_4 = m - \frac{\epsilon_1 + \epsilon_2}{2}.
\end{equation}
Note that the condition $\epsilon_1 + \epsilon_2 + \epsilon_3 + \epsilon_4 = 0$ is automatically satisfied. The elliptic genus of $n$ M-strings, with generic fugacities, was computed in \cite{Haghighat:2013gba}; we denote this elliptic genus by
\begin{equation}
	Z^{\textrm{M-str}}_n(\tau, \epsilon_1, \epsilon_2, m).
\end{equation}
In the next section we will review its computation and its connection to the $\mathcal{N}=2^*$ partition function.

\section{Review of M-strings}
\label{sec:3}

M-strings arise naturally in the context of $A_{N-1}$ $(2,0)$ theories and capture the spectrum of BPS states which arise when deforming away from the CFT point \cite{Haghighat:2013gba}. It was shown there that the 5d BPS index one obtains from performing a compactification of these theories on $T^2$ with general twists can be written in terms of a sum of elliptic genera of different numbers of M-strings. In particular, for two M5 branes we have the relation
\begin{equation} \label{eq:Mstrpf}
	Z^{\textrm{5d}~\mathcal{N}=2^* SU(2)} = \sum_{n} Q^n Z_n^{\textrm{M-str}}(\tau,\epsilon_1,\epsilon_2,m),
\end{equation}
where 
\begin{equation}
	Q = e^{2\pi i t},
\end{equation}
$t$ being the Coulomb branch parameter of the gauge theory. Since the Coulomb branch parameter is the separation between the M5 branes along an interval $I$, one sees that $t$ is also the tension of the self-dual strings of the $(2,0)$ theory, \textit{i.e.}\! the M-strings. In the expansion (\ref{eq:Mstrpf}) the M2 branes wrapped on $T^2 \times I$ play the role of instantons whose moduli space gives rise to a path integral representation of the elliptic genus \cite{Haghighat:2013tka,Haghighat:2013gba}. This leads to the computation of the partition function of the 5d $ \mathcal{N}=2^* $ $ SU(2) $ theory in terms of the elliptic genera of M-strings.\\

\noindent Let us next review how $Z_n^{\textrm{M-str}}$ is computed. As shown in \cite{Haghighat:2013gba} it can be decomposed into ``domain wall'' contributions as follows
\begin{equation}
	Z_n^{\textrm{M-str}} = \sum_{|\nu|=n} D^{\textrm{M5}}_{\emptyset \nu} D^{\textrm{M5}}_{\nu \emptyset},
\end{equation}
where $\nu$ are Young tableaux with $n$ boxes and $\emptyset$ denotes the empty Young tableau. The $D^{\textrm{M5}}_{\mu \nu}$ factors can be interpreted as domain walls of ABJM theory which interpolate between two different vacua. More precisely, it is known that ABJM theory \cite{Aharony:2008ug} on $T^2\times \mathbb{R}$ has vacua labeled by Young tableaux \cite{Kim:2010mr}. Thus reducing the theory on $T^2$ gives rise to a quantum mechanics whose Hilbert space is labelled by Young tableaux, where the euclidean time flows along $X^6$. We can then define the $D^{\textrm{M5}}$ as matrix elements in this quantum mechanics with an insertion of an M5 brane defect operator as follows:
\begin{equation}
	D^{\textrm{M5}}_{\mu \nu} = \langle \nu | \widehat{D}^{\textrm{M5}} | \mu \rangle.
\end{equation}

\begin{figure}[here!]
  \centering
	\includegraphics[width=0.6\textwidth]{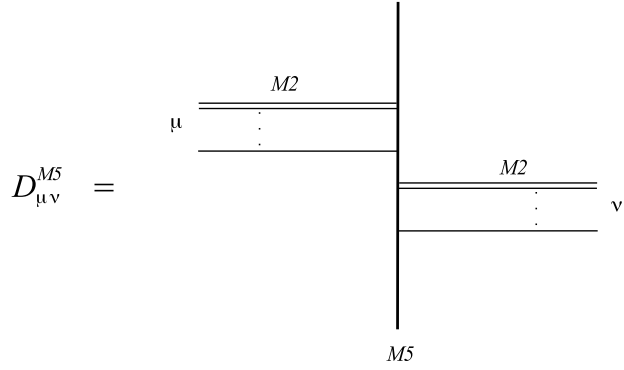}
  \caption{A M2 domain wall with a M5 defect.}
  \label{fig:DM5}
\end{figure}

A pictorial representation of this operator is shown in Figure \ref{fig:DM5}.

\noindent This operator was computed in \cite{Haghighat:2013gba} by means of the refined topological vertex \cite{Iqbal:2007ii}, and the result is given by
\begin{align}\nn
&D^{\textrm{M5}}_{\nu\mu}(\tau,m,\epsilon_1,\epsilon_2)=t^{-\frac{\Arrowvert\mu^{t}\Arrowvert^2}{2}}\,q^{-\frac{\Arrowvert\nu\Arrowvert^2}{2}}Q_{m}^{-\frac{|\nu|+|\mu|}{2}}\\ \nn &\times\prod_{k=1}^{\infty}\prod_{(i,j)\in\nu}\frac{(1-Q_{\tau}^{k}Q_{m}^{-1}\,q^{-\nu_{i}+j-\frac{1}{2}}\,t^{-\mu_{j}^{t}+i-\frac{1}{2}})
(1-Q_{\tau}^{k-1}Q_{m}\,q^{\nu_{i}-j+\frac{1}{2}}\,t^{\mu_{j}^{t}-i+\frac{1}{2}})}{(1-Q_{\tau}^{k}\,q^{\nu_{i}-j}\, t^{\nu_{,j}^{t}-i+1})(1-Q_{\tau}^{k-1}\,q^{-\nu_{i}+j-1}\,t^{-\nu_{j}^{t}+i})}\\
&\times\prod_{(i,j)\in\mu}\frac{(1-Q_{\tau}^{k}Q_{m}^{-1}\,q^{\mu_{i}-j+\frac{1}{2}}t^{\nu_{j}^{t}-i+\frac{1}{2}})(1-Q_{\tau}^{k-1}Q_{m} \,q^{-\mu_{i}+j-\frac{1}{2}}t^{-\nu_{j}^{t}+i-\frac{1}{2}})}{(1-Q_{\tau}^{k}\,q^{\mu_{i}-j+1}t^{\mu_{j}^{t}-i})(1-Q_{\tau}^{k-1}\,q^{-\mu_{i}+j}t^{-\mu_{j}^{t}+i-1})}, \label{eq:domainwall}
\end{align}
where we have defined
\begin{equation}
	q = e^{2\pi i \epsilon_1}, \quad t = e^{-2\pi i \epsilon_2}, \quad Q_m = e^{2\pi i m}, \quad Q_{\tau} = e^{2\pi i \tau}.
\end{equation}
We now specialize to the cases of interest for us, namely $D^{\textrm{M5}}_{\emptyset \nu}$ and $D^{\textrm{M5}}_{\nu \emptyset}$, and introduce the notation
\begin{equation}
	\xi_+(\tau;z) = \prod_{k\geq 1} (1-Q_{\tau}^k e^{2\pi i z}), \quad \xi_-(\tau;z) = \prod_{k \geq 1}(1-Q_{\tau}^{k-1} e^{-2\pi i z}).
\end{equation}
The functions $\xi_-(\tau; z)$ and $\xi_+(\tau;z)$ are quantum dilogarithms which can be thought of as ``half theta functions'', as they combine nicely into a theta function
\begin{equation}
	-i e^{-i \pi z} e^{\frac{\pi i \tau}{6}}\eta(\tau) \xi_-(\tau;z) \xi_+(\tau;z) = \theta_1(\tau;z),
\end{equation}
so that the product $D^{\textrm{M5}}_{\emptyset \nu} D^{\textrm{M5}}_{\nu \emptyset}$ is a modular function in $\tau$.\\

\noindent Up to prefactors, one finds\footnote{We will ignore prefactors here as well as in the rest of the paper, since once two domain walls are glued together all surviving prefactors can be removed by a redefinition of $ Q $, as in \cite{Haghighat:2013gba}.}
\begin{eqnarray}
	D^{\textrm{M5}}_{\emptyset \nu} & = & \prod_{(i,j)\in\nu}\frac{\theta_1(\tau;-m+\epsilon_1(\nu_i - j +1/2) - \epsilon_2(-i+1/2))\eta(\tau)^{-1}}{\xi_-(\tau;\epsilon_1(\nu_i-j) - \epsilon_2(\nu^t_j - i +1))\xi_+(\tau;\epsilon_1(\nu_i-j+1) - \epsilon_2(\nu^t_j-i))},\hspace{.3in} \\
	D^{\textrm{M5}}_{\nu \emptyset} & = & \prod_{(i,j)\in\nu}\frac{\theta_1(\tau;-m-\epsilon_1(\nu_i-j+1/2)+\epsilon_2(-i+1/2))\eta(\tau)^{-1}}{\xi_-(\tau;\epsilon_1(\nu_i-j+1)-\epsilon_2(\nu^t_j-i))\xi_+(\tau;\epsilon_1(\nu_i-j)-\epsilon_2(\nu^t_j-i+1))}.\hspace{.3in}\label{eq:DM5}
\end{eqnarray}
Note that $D^{\textrm{M5}}_{\emptyset \nu}$ and $D^{\textrm{M5}}_{\nu \emptyset}$ get exchanged under the map\footnote{This also leads to an overall $ (-1)^{|\nu|}  $ factor multiplying $ D_{\emptyset \nu} D_{\nu \emptyset} $ which can always be absorbed by shifting $ Q \to -Q $.}
\begin{equation}
	m \mapsto -m, \quad \xi_{\pm} \mapsto \xi_{\mp}.\label{eq:leftright}
\end{equation}

Indeed one can immediately see, using the above building blocks, that the partition function of two M5 branes (\ref{eq:Mstrpf}) has the following form:
\begin{eqnarray} \label{eq:Mstrpfres}
	Z^{5d~\mathcal{N}=2^* SU(2)} & = & \sum_{\nu} Q^{|\nu|} \prod_{(i,j)\in \nu} \frac{\theta_1(\tau;z_{ij}) \theta_1(\tau;v_{ij})}{\theta_1(\tau;w_{ij})\theta_1(\tau;u_{ij})},
\end{eqnarray}
where following \cite{Haghighat:2013gba} we have defined
\begin{eqnarray}
	e^{2\pi i z_{ij}} = Q_m^{-1} q^{\nu_i - j +1/2} t^{-i+1/2}, & \quad & e^{2\pi i v_{ij}} = Q_m^{-1} t^{i-1/2} q^{-\nu_i+j-1/2},  \nonumber\\
	e^{2\pi i w_{ij}} = q^{\nu_i -j +1} t^{\nu_j^t -i}, & \quad & e^{2\pi i u_{ij}}  =  q^{\nu_i -j} t^{\nu^t_j - i +1}.
\end{eqnarray}
One can clearly see from the expression in (\ref{eq:Mstrpfres}) that the elliptic genus of $n$ M-strings receives contributions from $4n$ bosons as well as from $4n$ fermions coming from the theta functions in the denominator and numerator respectively. These have the interpretation of coordinates on the target space which is the moduli space of $n$ $U(1)$ instantons on $\mathbb{R}^4$ in the sigma model description of M-strings \cite{Haghighat:2013gba}. The elliptic genus can be computed by localization on the target space, which in this case is the Hilbert scheme of $n$ points on $\mathbb{C}^2$, namely $\textrm{Hilb}^n[\mathbb{C}^2]$. Localization is done with respect to a $U(1)^2$ action with generators $\epsilon_1$ and $\epsilon_2$ and the path integral turns into a sum over the fixed points of this action which are labelled by Young tableaux. The coefficients of $\epsilon_1$ and $\epsilon_2$ in the theta functions in the numerator are the weights of the $U(1)^2$ action on the fermions while those in the denominator are the corresponding ones for the bosons. The different weights reflect the fact that, while the bosons are sections of the tangent bundle, the right-moving fermions transform as sections of the tautological bundle and therefore supersymmetry in the right-moving sector is broken.\\

\noindent We would also like to remark here that the elliptic genera $Z^{\textrm{M-str}}_n$ satisfy a holomorphic anomaly equation derived in \cite{Haghighat:2013gba}. To see this note that the sum in (\ref{eq:Mstrpfres}) is not modular invariant, as under $SL(2,\mathbb{Z})$ transformations each summand transforms with a different phase factor:
\begin{equation} \label{eq:modanomaly}
	Z_n^{\textrm{M-str}}(-\frac{1}{\tau},\frac{\epsilon_1}{\tau},\frac{\epsilon_2}{\tau},\frac{m}{\tau}) = e^{\frac{2\pi i}{\tau} (\epsilon_1 \epsilon_2 n^2 + (m^2 - (\epsilon_+/2)^2)n)} Z^{\textrm{M-str}}_n(\tau,\epsilon_1,\epsilon_2,m).
\end{equation}
To compensate for this phase factor and make the full partition function a modular function, one needs to make the theta function nonholomorphic. This is done by using its expansion in terms of Eisenstein series
\begin{equation}
	\theta_1(\tau;z) = \eta(\tau)^3 (2\pi z) \exp\left(\sum_{k\geq 1} \frac{B_{2k}}{(2k)(2k)!} E_{2k}(\tau)(2\pi iz)^{2k}\right),
\end{equation}
and making the replacement
\begin{equation} \label{eq:E2hat}
	E_2(\tau) \rightarrow \widehat{E}_2(\tau,\bar{\tau}) \equiv E_2(\tau) - \frac{3}{\pi \textrm{Im}(\tau)}.
\end{equation}
From now on we will often suppress the dependence on $\tau$ in the modular forms we use, as well as in $\xi_{\pm}$.
Using these modified theta functions one can check easily using (\ref{eq:Mstrpfres}) that the elliptic genus of $n$ M-strings, which is no longer holomorphic, satisfies the following holomorphic anomaly equation:
\begin{equation} 
	\frac{\partial Z^{\textrm{M-str}}_n}{\partial \widehat{E}_2} = -\frac{(2\pi)^2}{12}\left(\epsilon_1 \epsilon_2 n^2 + (m^2 - (\epsilon_+/2)^2) n\right) Z^{\textrm{M-str}}_n,\label{eq:MStrAnomaly}
\end{equation}
where $\epsilon_+ = \epsilon_1 + \epsilon_2$.  Said differently, we are trading here the modular anomaly of (\ref{eq:modanomaly}) with the holomorphic anomaly of (\ref{eq:MStrAnomaly}). More generally, whenever we encounter in this paper a Jacobi form with a modular anomaly
\begin{equation}
	Z_n(-\frac{1}{\tau},\frac{z_1}{\tau},\cdots,\frac{z_k}{\tau}) = e^{\frac{\pi i}{\tau} \alpha_n(z_1,\cdots,z_k)} Z_n(\tau,z_1,\cdots,z_k),
\end{equation}
we replace it with a non-holomorphic but modular Jacobi form with a holomorphic anomaly\footnote{We know that this statement holds for the single variable holomorphic Jacobi forms we consider here, and we also expect it to be true for the class of meromorphic or multivariate Jacobi forms that are used in this paper.} 
\begin{equation}
	\frac{\partial Z_n(\tau,\bar{\tau},z_1,\cdots,z_k)}{\partial \widehat{E}_2} = -\frac{(2\pi)^2}{24} \alpha_n(z_1,\cdots,z_k) Z_n(\tau,\bar{\tau},z_1,\cdots,z_k).
\end{equation}
Thus the concepts modular and holomorphic anomaly are interchangable and whenever we are talking about one of them one should keep in mind that an analogous statement holds for the other.
\\

\begin{figure}[here!]
  \centering
	\includegraphics[width=0.6\textwidth]{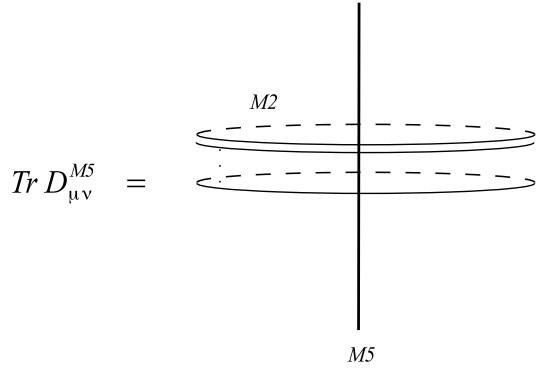}
  \caption{A M2 domain wall with a M5 defect where the $X^6$ direction is taken to be circular.}
  \label{fig:TrDM5}
\end{figure}

\noindent It is instructive to pause here and consider a slight modification of the above M-string setup.  To this end we look at a geometry which arises by taking the trace of a single domain wall as shown in Figure \ref{fig:TrDM5}.  
This configuration describes a six-dimensional supersymmetric $U(1)$ gauge theory \cite{Hollowood:2003cv} with an adjoint hypermultiplet of mass $m$ whose partition function is given by a sum over elliptic genera which correspond to the domain-wall traces as follows
\begin{equation} \label{eq:sixdpf}
	Z^{6d~U(1)} = \sum_n Q^n \sum_{|\nu|=n} D^{M5}_{\nu \nu}.
\end{equation}
The fundamental objects of this theory are known as ``little strings'' \cite{Seiberg:1997zk}. From the point of view of the six-dimensional $U(1)$ gauge theory which arises in the weak coupling limit of type IIA string theory the string is a solitonic object whose moduli space is that of $U(1)$ instantons. Note that in the compactification of the six-dimensional theory on $T^2$ the string wrapping the torus becomes a $U(1)$ gauge instanton in four dimensions. Using the explicit definition of the M5 domain wall formula (\ref{eq:domainwall}) we obtain
\begin{eqnarray}\label{eq:symstrgenus}
	\sum_{|\nu|=n} D^{M5}_{\nu \nu} & = & \prod_{(i,j) \in \nu} \frac{\theta_1(\tilde{m} + \epsilon_1(\nu_i -j )-\epsilon_2(\nu_j^t -i + 1))}{\theta_1(\epsilon_1(\nu_i-j)-\epsilon_2(\nu_j^t-i+1))} \nonumber \\
	~ & ~ & \times \frac{\theta_1(-\tilde{m}+\epsilon_1(\nu_i-j+1)-\epsilon_2(\nu_j^t-i+1))}{\theta_1(\epsilon_1(\nu_i-j+1)-\epsilon_2(\nu_j^t-i))},
\end{eqnarray}
where we have redefined the R-symmetry generator $U(1)_m$ to be
\begin{equation}
	\tilde{m} = m + \frac{\epsilon_1 - \epsilon_2}{2}.
\end{equation}
Analogously to the M-string case, the expression (\ref{eq:symstrgenus}) can be interpreted as the elliptic genus of a sigma model with target space $\textrm{Hilb}^n[\mathbb{C}^2]$. But while in (\ref{eq:Mstrpfres}) the fermions and bosons transformed as sections of different bundles one can clearly see upon inspection of the $U(1)^2$ weights that in (\ref{eq:symstrgenus}) they both are sections of the tangent bundle. Therefore, supersymmetry in the right-moving sector is unbroken. Remarkably, in contrast to the M-string setup, the partition function (\ref{eq:sixdpf}) has another representation which is fundamentally different from the one given in (\ref{eq:symstrgenus}). This fact was first noticed for the unrefined case in \cite{Hollowood:2003cv} where the authors rewrote the result in terms of the symmetric product elliptic genus of \cite{Dijkgraaf:1996xw}. The underlying reason for this is the equivalence of the Hilbert scheme of points with the resolution of the singular space of the $n$-fold symmetric product of $\mathbb{R}^4$:
\begin{equation}
	\textrm{Hilb}^n[\mathbb{C}^2] = \textrm{Res}\left(\textrm{Sym}^n (\mathbb{R}^4)\right).
\end{equation}
Using instead of the Hilbert scheme the orbifold $\textrm{Sym}^n(\mathbb{R}^4)$ as the target space of the sigma model one arrives at an equivalent formula for the partition function (\ref{eq:sixdpf}):
\begin{equation} \label{eq:symprod}
	Z^{6d~U(1)} = \sum_{n=1}^{\infty} Q^n \chi(\textrm{Sym}^n(\mathbb{R}^4)) = \prod_{n=1,k=0}^{\infty} \prod_{p1,p2,p3} \frac{1}{(1-Q^n Q_{\tau}^k q^{p1} t^{p2} Q_{\tilde{m}}^{p3})^{c(kn,p1,p2,p3)}},
\end{equation}
 where in the above $\chi(\textrm{Sym}^n(\mathbb{R}^4)$ is the elliptic genus of $n$ strings and the expansion of the elliptic genus of one string is taken to be
 \begin{equation}
 	\chi(\mathbb{R}^4) = \sum_{k\geq 0, p1,p2,p3} c(k,p1,p2,p3) Q_{\tau}^k q^{p1} t^{p2} Q_{\tilde{m}}^{p3}.
\end{equation}
Thus in some sense the $n$-string result is fully determined in terms of the $1$-string result, which is a reflection of the fact that the $n$-string sector is obtained by winding single strings multiple times around the different cycles of $ T^2 $. It is remarkable that such a fundamentally different representation of the elliptic genus exists as the individual terms appearing in the two expansions (\ref{eq:sixdpf}) and (\ref{eq:symprod}) have a completely different pole structure in $\epsilon_1$, $\epsilon_2$ and $m$ as can be seen from (\ref{eq:symstrgenus}). Note that in the case of the $ A_1 $ (2,0) theory a symmetric product representation for the M-string elliptic genus does not exist; this hints at the existence of bound states of M-strings.
\section{Heterotic strings from orbifolding}
\label{sec:4}

Let us recall in this section the basics of the $ E_8\times E_8 $ heterotic string for the geometry considered in Section \ref{sec:2}. To start, we note that the Hilbert space of $n$ heterotic strings wrapping the $T^2$ is the symmetric product of the Hilbert space of a single heterotic string, as heterotic strings do not form bound states. Said differently, at the level of the free energy the $n$-heterotic string result is the same as the $n$-times wound single heterotic string. This can be used to compute the $Z^{\textrm{Het}}_n$ purely from the knowledge of $Z^{\textrm{Het}}_1$. To proceed we thus just have to know the result for one heterotic string, which we discuss next. Henceforth we will be working in lightcone gauge. For generic twist parameters $\epsilon_i,\, i=1,\cdots,4$, we have $(2,0)$ SUSY on the worldsheet. Thus there are four chiral multiplets with twisted boundary conditions coming from $\mathbb{R}^4_{2345}$ and $\mathbb{R}^4_{78910}$. When computing the elliptic genus the supersymmetric side contributes just a factor of $1$, as bosonic and fermionic degrees of freedom cancel out, but the non-supersymmetric side depends on all $8$ spacetime bosons which organize themselves into $4$ complex bosons with twisted boundary conditions:
\begin{equation} \label{eq:hetbosons}
	Z_{\textrm{bosons}}(\tau,\epsilon_i) = -\frac{\eta^4}{\theta(\epsilon_1)\theta(\epsilon_2)\theta(\epsilon_3)\theta(\epsilon_4)}.
\end{equation}
Furthermore, as the string is charged under the $E_8 \times E_8$ current algebra there will be also a bosonic path integral which contributes a factor of the character of $E_8 \times E_8$:
\begin{equation} \label{eq:E8char}
	\chi_{E_8 \times E_8} = \frac{\Theta_{E_8}(\tau;\vec{m}_{E_8,L})\Theta_{E_8}(\tau;\vec{m}_{E_8,R})}{\eta^{16}},
\end{equation}
where we have introduced the $E_8$-Weyl invariant theta function of modular weight $4$ and level $1$
\begin{equation}
	\Theta_{E_8}(\tau;\vec{m}) = \frac{1}{2} \sum_{i=1}^4 \theta_i(\tau;\vec{m})^8.
\end{equation}
Combining the factor (\ref{eq:E8char}) with the contributions from the $8$ spacetime bosons (\ref{eq:hetbosons}) one arrives at:
\begin{equation} \label{eq:1hetstring}
	Z^{\textrm{Het}}_1 = -\frac{\Theta_{E_8}(\vec{m}_{E_8,L})\Theta_{E_8}(\vec{m}_{E_8,R})}{\eta^{12} \theta_1(\epsilon_1) \theta_1(\epsilon_2)\theta_1(\epsilon_3)\theta_1(\epsilon_4)}.
\end{equation}
To obtain the formula for the elliptic genus of $n$ heterotic strings we can apply the results of \cite{Dijkgraaf:1996xw}. First of all, we expect the full partition function of heterotic strings to be:
\begin{equation} \label{Hetpf}
	Z^{\textrm{Het}} = \sum_{n \geq 0} Q^n Z^{\textrm{Het}}_n,
\end{equation}
where $Q = e^{2\pi i \rho}$ with $\rho$ being the complexified K\"ahler parameter of the $T^2$ and $Z^{\textrm{Het}}_0$ is taken to be $1$. We can next obtain the one-loop free energy\footnote{This free energy can also be computed through a heterotic string one-loop amplitude, hence the name. Similar computations have for example been performed in \cite{Marino:1998pg}.} by taking the logarithm of the partition function
\begin{equation}
	\mathcal{F}^{\textrm{Het}} = \log(Z^{\textrm{Het}}) = \sum_{n \geq 1} Q^n \mathcal{F}^{\textrm{Het}}_n.
\end{equation}
Now, following \cite{Dijkgraaf:1996xw} we can express $\mathcal{F}^{\textrm{Het}}_n$ in terms of $\mathcal{F}^{\textrm{Het}}_1$ through the Hecke transform:
\begin{equation}
	\mathcal{F}^{\textrm{Het}}_n = T_n \mathcal{F}^{\textrm{Het}}_1,
\end{equation}
where the Hecke operator $T_n$ acts on a weak Jacobi form $f(\tau,z)$ of weight $k$ as
\begin{equation}
	T_n f(\tau,z) = n^{k-1} \sum_{\stackrel{ad=n}{a,d>0}} \frac{1}{d^k} \sum_{b~ (\textrm{mod} ~d)} f\left(\frac{a\tau+b}{d},a z\right).
\end{equation}
Applying this to our setup and noting that $\mathcal{F}^{\textrm{Het}}_1 = Z^{\textrm{Het}}_1$ has modular weight zero we obtain
\begin{equation} \label{eq:Hetpfres}
	Z^{\textrm{Het}} = \exp\left[\sum_{n\geq 0} Q^n \frac{1}{n} \sum_{\stackrel{ad=n}{a,d>0}}\sum_{b(\textrm{mod}~d)} Z_1^{\textrm{Het}}\left(\frac{a \tau+b}{d},a \epsilon_i, a \vec{m}\right) \right],
\end{equation}
which together with (\ref{Hetpf}) allows us to compute $Z^{\textrm{Het}}_n$. Again, in order for (\ref{eq:Hetpfres}) to be modular, analogously to the M-string case one has to introduce some non-holomorphicity; the resulting holomorphic anomaly can be deduced from the modular anomaly. In the case of a single string this anomaly can be read off from (\ref{eq:1hetstring}):
\begin{equation}
	Z^{\textrm{Het}}_1(-\frac{1}{\tau},\frac{\vec{\epsilon}}{\tau},\frac{\vec{m}}{\tau}) = \exp\left[-\frac{\pi i}{\tau} (\sum_{i=1}^4 \epsilon_i^2 - \sum_{i=1}^{16} m_i^2)\right] Z^{\textrm{Het}}_1(\tau,\vec{\epsilon},\vec{m}),
\end{equation}
which shows that $Z^{\textrm{Het}}_1$ is a weight zero Jacobi form of index $1/2$ in each of the elliptic parameters $\epsilon_i$ and $m_i$. As the order $n$ Hecke operator transforms an index $m$ Jacobi form to an index $nm$ Jacobi form we see that the elliptic genus for $n$ heterotic strings suffers from the anomaly
\begin{equation}
	Z^{\textrm{Het}}_n(-\frac{1}{\tau},\frac{\vec{\epsilon}}{\tau},\frac{\vec{m}}{\tau}) = \exp\left[-\frac{\pi i}{\tau} n(\sum_{i=1}^4 \epsilon_i^2 - \sum_{i=1}^{16} m_i^2)\right] Z^{\textrm{Het}}_n(\tau,\vec{\epsilon},\vec{m}).\label{eq:hetanomaly}
\end{equation}

\section{Review of known results for E-strings}
\label{sec:5}

In this section we recall a geometric setup which gives rise to the E-string theory, as well as topological string computations on this geometry that are related to the computation of the E-string elliptic genus. To do so, we first start with the F-theory realization of the six-simensional superconformal field theory whose degrees of freedom are the E-string, as well as its M-theory dual. In a second subsection, we review the connection between the M-theory picture and topological strings and the way this connection has been exploited to compute the E-string free energy as a genus expansion.

\subsection{M- and F-theory realizations}
\label{sec:5a}

E-strings arise in the Coulomb branch of small instantons \cite{Witten:1995gx} in $E_8 \times E_8$ heterotic string compactifications on K3 \cite{Seiberg:1996vs,Ganor:1996mu}. In order to connect this to the picture of M2 branes suspended between M9 and M5 branes discussed in section \ref{sec:2} one embeds the small instanton in one of the $E_8$ factors and considers a specific limit of the K3 where its volume is sent to infinity and one zooms into the neighbourhood of the small instanton. In this limit the K3 can be replaced locally by $\mathbb{R}^4_{78910}$ with an M5 brane sitting at the origin of $\mathbb{R}^4_{78910}$ and wrapping $T^2 \times \mathbb{R}^4_{2345}$. Furthermore, the gauge fields of the two $E_8$ factors are on two different ``end of the world'' M9 planes as discussed in Section \ref{sec:2}. Moving the M5 brane away from the M9 plane one gains a tensor multiplet whose scalar component parametrizes the distance and hence this phase can be interpreted as Coulomb branch. On the other hand the phase where one lets the size of the instanton grow is the Higgs branch. 

These transitions have a beautiful F-theory realization obtained by compactifying F-theory on a elliptic Calabi-Yau threefold \cite{Witten:1996qb,Morrison:1996na,Morrison:1996pp}. Introducing a tensor multiplet by moving an M5 brane away from an M9 plane translates to blowing up the base of the elliptic fibration at a point. This leads locally to the replacement of $\mathbb{C}^2$ with its blow-up which can be identified with the bundle $\mathcal{O}(-1) \rightarrow \mathbb{P}^1$. In order for the Calabi-Yau property to be satisfied the elliptic fibration over the resulting $\mathbb{P}^1$ is chosen to be such that the resulting elliptic rational surface is the so called ``half-K3'' surface. Alternatively, this surface can also be described as the del Pezzo $9$ surface $\mathbb{B}_9$ obtained by blowing up $\mathbb{P}^2$ at $9$ points. The Calabi-Yau is then locally the anti-canonical bundle over this surface, namely
\begin{equation}
	\textrm{CY}_3 = \mathcal{O}(-K) \rightarrow \half K3.
\end{equation} 
In this picture the exceptional strings, which in the M-theory setup come from M2 branes suspended between the M9 and the M5 brane, arise from D3 branes wrapping the base of $\half K3$. These are pierced by $8$ 7-branes, corresponding to the deformation moduli of the elliptic fibration, and are thus expected to be charged under a $E_8$ current algebra. \\

\noindent Next, we employ the duality between F-theory and M-theory. This corresponds to compactifying our F-theory setup on $S^1$ to five dimensions, which is dual to M-theory on the $\half K3$ Calabi-Yau manifold. The D3 branes wrapping $n$ times the base of $\half K3$ and having KK-momentum $k$ along $S^1$ map in the dual picture to M2 branes which again wrap the base of $\half K3$ $n$ times but are now also wrapping the elliptic fiber $k$ times. In the following section we utilize the topological string to compute the BPS degeneracies associated with these states.

\subsection{Results from the topological string on $ \frac{1}{2} $-K3}
\label{sec:5b}

Let us now review the connection between the M-theory setup we have arrived at and the refined topological string, and how the latter can be used to compute the elliptic genus of $ n $ E-strings. We have arrived through various dualities at the following setup:
\[ \text{M-theory on }\; (S^1 \times \mathbb{C}^2)_{\epsilon_1,\epsilon_2} \times X, \]
where
\begin{equation}
X = \mathcal{O}(-K) \to \frac{1}{2}K3
\end{equation}
is the local Calabi-Yau threefold given by the anti-canonical bundle over the half-K3 surface discussed above. In going around the circle, one twists the two copies of $ \mathbb{C} $ in $ \mathbb{C}^2 $ respectively by $ \epsilon_1 $ and $ \epsilon_2 $, and furthermore performs a rotation of the fiber of $ X $ by $ -\frac{\epsilon_1+\epsilon_2}{2} $ in order to preserve supersymmetry. In this setup, one can count the number of BPS configurations of M2-branes wrapping cycles in $ X $. These are precisely the states that are counted by the A-model refined topological string partition function on $ X $ \cite{Dijkgraaf:2006um}, so the following statement holds:
\begin{equation}
Z_{M-theory}(S^1\times \mathbb{C}^2_{\epsilon_1,\epsilon_2}\times X) \equiv Z_{top}(X;\epsilon_1,\epsilon_2).
\end{equation}
Besides depending on $ \epsilon_1, \epsilon_2 $, the topological string partition function also depends on the K\"ahler parameters associated to the two-cycles of $ X $. The second-degree homology of the local half-K3 Calabi-Yau is given by
\begin{equation}
H_2(X,\mathbb{Z}) = \Gamma^{E_8} \oplus \Gamma_{1,1},
\end{equation}
where $ \Gamma_{1,1} $ is the two-dimensional hyperbolic lattice generated by the $ \mathbb{P}^1 $ base of $\mathbb{B}_9 $, of area $ t $, and the torus fiber, of area $ \tau $; $ \Gamma^{E_8} $, on the other hand, is the $ E_8 $ lattice generated by eight additional two-cycles of area $ (m_{E_8,1},\dots,m_{E_8,8}) $. Therefore the topological string partition function for this geometry is a function of the 12 parameters $ (\epsilon_1,\epsilon_2,t,\tau,\vec{m}_{E_8}) $.\\

\noindent At the same time, upon compactification on X, one obtains an effective 5d gauge theory on $ \mathbb{C}^2\times S^1 $, where $ S^1 $ plays the role of the thermal circle. This is the $ Sp(1)\approx SU(2)$ gauge theory with eight fundamental hypermultiplets of \cite{Douglas:1996xp,Ganor:1996pc,Seiberg:1996bd}, obtained from the worldvolume theory of the M5 brane in the presence of the M9 plane by first reducing along a circle to obtain a 5d theory, and then further compactifying along the thermal circle. The theory has a superconformal fixed point at strong coupling where the flavor group is enhanced to affine $ E_8 $. The BPS configurations of M2 branes that are counted by the topological string partition function give rise to BPS particles of this gauge theory.  From the point of view of the 5d gauge theory, $ (\epsilon_1,\epsilon_2) $ are fugacities associated to the $ U(1)\times U(1) $ Cartan subgroup of the little group $ SO(4) = SU(2)\times SU(2) $. Furthermore, $ t $ corresponds to the Coulomb branch parameter of the theory (which descends from the vev of the 6d tensor multiplet parametrizing the separation between the M5 branes and M9 planes); $ \tau $ is related to the 5d gauge coupling as follows: 
\begin{equation}
\tau = \frac{2\pi i}{g_{YM}^2};
\end{equation}
finally, $ \vec{m}_{E_8} $ are simply the masses of the eight hypermultiplets.\\

\noindent Since the gauge theory has its origin from the 6d theory of the M5 brane, the BPS instantons of the gauge theory are in one-to-one correspondence with the states of the E-string wrapping the torus. One is thus led to the following relation between the refined topological string (i.e. the 5d BPS index) and the E-string elliptic genus:
\begin{equation}
Z_{top}(X; \epsilon_1,\epsilon_2) = \sum_{n=0}^\infty Q_{t}^{\,n} Z^{E-str}_{n}(\tau;\epsilon_1,\epsilon_2,\vec{m}_{E_8});
\end{equation}
that is, the coefficient of $ Q^{n} = \exp(2\pi i\, n\, t) $, where $ t $ is interpreted as the string tension, counts the states coming from $ n $ E-strings wrapping the torus\footnote{Note that we take this sector to include both a single string wrapping the torus $ n $ times and configurations of $ k $ strings wrapping the torus respectively $ n_1,\dots,n_k $ times, such that $ \sum_{j=1}^k n_k = n $.} . It is by exploiting this connection with topological strings that it has been possible to perform explicit computations of the E-string elliptic genus, beginning with the work of \cite{Klemm:1996hh} in the context of unrefined topological strings (i.e. setting $ \epsilon_2 = -\epsilon_1 = g_s $). More precisely, in this context one aims at computing the topological string free energy as a perturbative expansion which takes the following form:
\begin{equation}
\mathcal{F} \equiv  \log\left(Z_{top}(X;\epsilon_1,\epsilon_2)\right) = \sum_{n\geq 0}\sum_{g\geq 0} Q^{n} g_s^{2g-2}\mathcal{F}_{n,\,g}.
\end{equation}
The free energy of a single E-string is known to arbitrary genus (see the discussion in Section \ref{sec:6c}); in the case of several strings, topological string techniques have been employed to compute the free energy to high genus (for instance, in \cite{Mohri:2001zz} the free energy of up to five E-strings is computed up to $ g = 5 $). Recently, a similar approach was successfully developed \cite{Huang:2013yta} in the refined case (where $ \epsilon_1,\epsilon_2 $ are taken to be arbitrary), generalizing techniques that were employed in the unrefined limit. In the refined case, the free energy takes the following form:
\begin{equation}
\mathcal{F} = \sum_{n \geq 0}\sum_{g\geq 0}\sum_{n \geq 0} Q^{n} (-\epsilon_1\epsilon_2)^{g-1}(\epsilon_1+\epsilon_2)^{2\ell}\mathcal{F}_{n,g,\ell}.
\end{equation}
One then observes that (in the case where $ \vec{m}_{E_8} $ are set to zero) the free energy satisfies the following modular anomaly equation (which immediately gives the holomorphic anomaly equation upon replacing $ E_2(\tau) $ with its modular completion $ \widehat{E}_{2}(\tau,\overline\tau) $):
\begin{eqnarray}
\partial_{E_2}\mathcal{F}_{n,g,\ell} &=& \frac{1}{24}\sum_{\nu = 1}^{n-1}\sum_{\gamma=0}^g \sum_{\lambda=0}^{\ell} \nu(n-\nu)\mathcal{F}_{\nu,\gamma,\lambda}\mathcal{F}_{n-\nu,g-\gamma,\ell-\lambda}\nonumber\\
&+& \frac{n(n+1)}{24}\mathcal{F}_{n,g-1,\ell}-\frac{n}{24}\mathcal{F}_{n,g,\ell-1}.\label{eq:modanomalyF}
\end{eqnarray}
This generalizes the modular anomaly equation that was found in the unrefined case by Hosono \textit{et al.} \cite{Hosono:1999qc}; the form of this expression was determined (up to the $ n/24 $ coefficient in front of the last term, which was obtained by other means) by requiring it to reduce correctly to known expressions in different limits. It is known that $n$ E-strings can form bound states and thus admit no simple description in terms of the Hecke transform of a single string as was the case for heterotic strings. This is reflected in the $\mathcal{F}_\nu \mathcal{F}_{n-\nu}$ term of the holomorphic anomaly equation (\ref{eq:modanomalyF}). As has been noted in \cite{Minahan:1998vr} (see also \cite{Alim:2010cf}) this term shows that bound states of $\nu$ and $n-\nu$ strings can pair up to form a configuration of $n$ E-strings. In Section \ref{sec:6} we will provide a simple new derivation of this formula using the M-theory realization of the E-string.\\

\noindent The modular anomaly equation allows one to fix the $ E_2$-dependent part of the  $ (n, g, \ell)  $ piece of the topological string free energy as long as the terms with lower values of $ n, g $ and $ \ell $ are known. It does not fix the $ E_2 $-independent piece; however, this is captured by a modular form of definite weight, and since the vector space of modular forms of a given weight is finitely generated, it can be uniquely determined by fixing a finite number of coefficients in the $ Q_\tau $ expansion of $ \mathcal{F}_{n,g,\ell} $. For low values of $ g $ and $ \ell $, these coefficients can be fixed by imposing `vanishing conditions', that is, the fact that certain contributions to the topological string free energy are required to vanish \cite{Huang:2013yta}. Unfortunately, it is known that as one increases the values of $ g $ and $ \ell $ the number of coefficients that need to be fixed grows faster than the number of vanishing conditions \cite{Huang:2013yta}, so one cannot use this method to compute the free energy to arbitrary order.\\

\noindent The same approach has been employed to compute the free energy for nonzero values of $ \vec{m}_{E_8} $ \cite{Huang:2013yta} (although again the vanishing conditions only allow one to determine the free energy for small enough values of $ \ell $ and $ g $). It is known \cite{Minahan:1998vr} that the contributions to the free energy coming from the $ n $-string sector can be written in terms of combinations of level $ n $ characters of the affine $E_8$ algebra; furthermore, these combinations of characters can be written as polynomials in nine Jacobi forms $ A_1, A_2, A_3,A_4, A_5, B_2,B_3, B_4, B_6, $\footnote{We refer the reader to Appendix \ref{sec:appendix} for more details on this class of Jacobi forms.} which are $ \vec{m}_{E_8} $-dependent and invariant under the Weyl group of $ E_8 $. The subscript in $ A_n, B_n $ indicates the amount by which they contribute to the $E_8$ level; for example, level 2 modular invariant combinations of characters of affine $ E_8 $ can be written as a linear combinations of $ A_1^2,A_2,B_2 $. Furthermore, $ A_{1,\dots, 5} $ are weight-4 Jacobi forms that reduce to the Eisenstein series $ E_4 $ in the limit $ \vec{m}_{E_8} \to 0 $, while $ B_{2,3,4,6} $ have weight 6 and reduce to $ E_6 $ in the same limit.  In the next section we will use the explicit results of \cite{Huang:2013yta} for the E-string free energy, written in terms of these Weyl[$ E_8 $]-invariant Jacobi forms, as input to compute the elliptic genus for two E-strings to arbitrary powers of $ \ell $ and $ g $.

\section{The elliptic genus of E- and heterotic Strings}
\label{sec:6}

In this section, we provide evidence that the elliptic genera of up to two heterotic and exceptional strings can be written in terms of domain wall contributions, analogously to the M-string case. Recall that, as discussed in Section \ref{sec:3}, the elliptic genus for $ n $ M-strings could be written in terms of M5 domain wall contributions as follows:
\begin{equation}
	Z_n^{\textrm{M-str}} = \sum_{|\nu|=n} D^{\textrm{M5}}_{\emptyset \nu} D^{\textrm{M5}}_{\nu \emptyset}.\label{eq:ZMstring}
\end{equation}
This result had a very natural interpretation from the point of view of the M2 branes suspended between the M5 branes along the $ X^6 $ direction. From the point of view of the M2 branes, the M5 branes are codimension one operators supported at a point along the $ X^6 $ direction. In the limit where the area of the $ T^2 $ is taken to be very small, one is left with one dimensional quantum mechanics, where $ X^6 $ plays the role of time; furthermore, the Hilbert space of $ n $ M2 branes is labeled by size $ n $ Young diagrams \cite{Haghighat:2013gba,Kim:2010mr}. These states are eigenfunctions of the Hamiltonian $ \widehat{\mathcal{H}} $, with eigenvalue given by the number of boxes in the Young diagram. Furthermore, the M5 defect operators become quantum mechanical operators that map a certain number of M2 branes to linear combinations of arbitrary numbers of M2 branes, so one can interpret the elliptic genus as expectation value of two M5 domain wall operators inserted at different times\footnote{Note that the definition of $ t $ we employ here differs by the one employed elsewhere by a Wick rotation, and therefore is rescaled by a factor of $ 2 \pi i $.}:
\[ e^{-n\, t} Z_n^{M-str} = \langle 0 | \widehat {D}^{M5}\, e^{-\widehat{\mathcal{H}} t} \widehat {D}^{M5}\, | 0\rangle. \]

\noindent Given that the difference between the E-string and the M-string is simply that in the former case the M2 branes terminate on an M9 plane, while in the latter they terminate on an M5 brane, it is natural to ask whether one can find a similar domain wall formula for the E-string elliptic genus, where instead of inserting an M5 domain wall operator on the left we take the product with an appropriate state $ \vert \psi_{M9}\rangle = D_{\nu}^{M9, L}\,\vert \nu \rangle $ for the M9 plane at the left end of $ S_1/\mathbb{Z}_2 $. In this section, therefore, we seek an expression for the E-string elliptic genus of the form
\begin{equation}
	e^{-n\, t}   Z_n^{\textrm{E-str}} = \sum_{|\nu|=n} \langle 0 | \widehat{D}^{M5} e^{- \widehat{H} t} \vert \nu \rangle\,\langle\nu\vert\psi_{M9}\rangle = \sum_{\vert\nu\vert = n} D^{\textrm{M9, L}}_{\nu} D^{\textrm{M5}}_{\nu \emptyset}.\label{eq:Estringdomain}
\end{equation}

\noindent Building on results from topological string theory and exploiting several properties that the elliptic genus of E-strings is expected to satisfy, we are able to uniquely fix the left M9 domain walls for one and two strings:
\begin{equation}
D^{M9, L}_{\ydiagram{1}} = \left(\frac{\Theta_{E_8}(\vec{m}_{E_8,L})}{\eta^8}\right)\frac{\eta}{\theta_1(\epsilon_3)}\cdot\frac{1}{\xi_-(\epsilon_1)\xi_+(-\epsilon_2)}
\end{equation}
and
\begin{align}
D_{\ydiagram{2}}^{M9,L} &= \frac{N_{\ydiagram{2}}(\vec{m}_{E_8,L},\epsilon_1,\epsilon_2)/\eta^{16}}{\xi_+(\epsilon_1)\xi_-(-\epsilon_2)\xi_-(\epsilon_1-\epsilon_2)\xi_+(2\epsilon_1)\theta_1(m+\epsilon_+/2)\theta_1(m+\epsilon_+/2+\epsilon_1)\eta^{-2}},\\
D_{\ydiagram{1,1}}^{M9,L} &=  \frac{N_{\ydiagram{1,1}}(\vec{m}_{E_8, L},\epsilon_1,\epsilon_2)/\eta^{16}}{\xi_-(\epsilon_1)\xi_+(-\epsilon_2)\xi_+(\epsilon_1-\epsilon_2)\xi_-(-2\epsilon_2)\theta_1(m-\epsilon_+/2)\theta_1(m-\epsilon_+/2-\epsilon_2)\eta^{-2}},
\end{align}
where $ N_{\ydiagram{2}} $ and $ N_{\ydiagram{1,1}}\, $ (explicit expressions for which can be found in Equations \eqref{eq:N2} and \eqref{eq:N11}) are certain Jacobi forms of weight 8 that depend on the $ E_8 $ Wilson lines as well as $ \epsilon_1,\epsilon_2 $.   We also find corresponding formulas for the right M9 domain walls. Combining these domain walls with the known M5 domain walls, we are able to reproduce the known elliptic genus for a single E-string, and find a novel closed formula for the two E-string elliptic genus, which takes the following form:
\begin{equation}
	Z_2^{\textrm{E-str}} = D^{\textrm{M9, L}}_{\ydiagram{2}} D^{\textrm{M5}}_{\ydiagram{2}~ \emptyset} + D^{\textrm{M9, L}}_{\ydiagram{1,1}} D^{\textrm{M5}}_{\ydiagram{1,1}~ \emptyset}.
\end{equation}

\noindent Once the M9 domain walls have been computed, it is natural to ask the following question: given that the heterotic string is given by M2 branes which end on two M9 planes, is it possible to also write the elliptic genus of heterotic strings in terms of domain wall expressions, where now we take both domain walls to be of the M9 type? We find that this is indeed the case: for one heterotic string, we show that
\begin{equation}
Z_1^{het} = D_{\ydiagram{1}}^{M9, L}(\vec{m}_{E_8, L})D_{\ydiagram{1}}^{M9, R}(\vec{m}_{E_8, R}).
\end{equation}
Similarly, for two heterotic strings we find the following identity:
\begin{equation}
Z_2^{het} = D_{\ydiagram{2}}^{M9, L}(\vec{m}_{E_8, L})D_{\ydiagram{2}}^{M9, R}(\vec{m}_{E_8, R}) + D_{\ydiagram{1,1}}^{M9, L}(\vec{m}_{E_8, L})D_{\ydiagram{1,1}}^{M9, R}(\vec{m}_{E_8, R})+(\dots),
\end{equation}
where $ (\dots) $ are two additional terms that are obtained from the first two by symmetrizing with respect to permutations of $ \epsilon_1, \epsilon_2, \epsilon_3 $ and $ \epsilon_4 $. This formula matches with the expression one obtains by using the Hecke transform of the one heterotic string elliptic genus, despite the fact that it has a very different appearance. In particular, each one of the terms appearing in our new expression is manifestly modular, and is split into two factors, one which only depends on the $ E_{8,L} $ degrees of freedom, and one which only depends on the $ E_{8,R} $ degrees of freedom.\\

\noindent While the computation of elliptic genera in this paper is limited to the case of one and two E-strings, we offer some evidence that our approach should work for an arbitrary number of strings by deriving the E-string modular anomaly equation, which was recently conjectured in \cite{Huang:2013yta}, from the known holomorphic anomaly equations for M- and heterotic strings, assuming that these elliptic genera can all be written in terms of M5 and M9 domain walls.\\

\noindent This section is divided as follows: in Section \ref{sec:6a} we provide more details of our approach and of the ingredients that go into it; in Section \ref{sec:6b} we show how this leads to an alternate derivation of the E-string modular anomaly equation. in Section \ref{sec:6c} we compute the M9 domain wall factor associated to a single M9 plane, and use it to reproduce known formulas for the elliptic genus for a single E-string and a single heterotic string. In \ref{sec:6d} we turn to the case of two strings; we derive expressions for the corresponding M9 domain walls, and we use these to derive a closed formula for the elliptic genus of two E-strings; furthermore, we obtain a novel expression for the elliptic genus of two heterotic strings. Finally, in \ref{sec:6e} we make some additional comments about the features of the domain wall expressions we obtained.

\subsection{M9 domain walls}
\label{sec:6a}

The computation of the elliptic genus of the E-strings is not an easy task, since configurations of several E-strings form bound states and therefore their elliptic genus cannot be deduced from the elliptic genus of a single E-string by means of the Hecke transform. Here we describe in some detail an alternative approach, which is based on the computation of M9 domain wall factors. Fortunately, from the symmetries of the problem one can deduce several properties that these factors must satisfy which will allow us to uniquely determine them in the case of $ n = 1, 2 $. We list the expected properties here:
\begin{itemize}
\item The elliptic genus for $ n $ E-strings is expected to transform with modular weight 0 under the $ SL(2,\mathbb{Z}) $ transformation 
\begin{equation}
(t,m,\epsilon_1,\epsilon_2,\tau) \quad \to \quad \left(t,\frac{m}{c\tau+d},\frac{\epsilon_1}{c\tau+d},\frac{\epsilon_2}{c\tau+d},\frac{a \tau + b}{c\tau + d}\right), 
\end{equation}
where $ \begin{pmatrix} a & b \\ c & d \end{pmatrix} \in SL(2,\mathbb{Z}) $.\\\newline Since the denominator of the M5 domain wall expression \eqref{eq:DM5} by itself is not modular invariant, each factor of $ \xi_{\pm}(z) $ that appears there must be matched by an equivalent factor of $ \xi_{\mp}(z) $ in the M9 domain wall in order to combine them into the Jacobi form $ \theta_1(\tau,z)/\eta(\tau) $, which has well-defined modular transformation properties.
\item The E-string partition function does not depend on the mass parameter $ m = (\epsilon_4-\epsilon_3)/2 $. This implies that the mass-dependent factors in the numerator of the M5 domain wall must be canceled by identical factors in the denominator of the M9 domain wall.

\item As discussed above, the $ \vec{m}_{E_8} $ dependence of the $ n $ E-string free energy (and therefore also the $ n $ E-string elliptic genus) is captured in terms of level $ n $ characters of affine $E_8$, and thus the $ \vec{m}_{E_8} $-dependent factors in the $ n $ E-string elliptic genus can be written in terms of level $ n $ combinations of the Weyl[$ E_8 $]-invariant Jacobi forms $ A_{1,2,3,4,5},B_{2,3,4,6} $ which are discussed in Appendix \ref{sec:appendix}.

\item From the modular anomaly equation for the E-string free energy, Eq. \eqref{eq:modanomalyF}, one easily derives a modular anomaly equation for the elliptic genus of $ n $ E-strings in the limit $ \vec{m}_{E_8} \to 0 $:
\begin{equation}
\partial_{E_2} Z_n^{E-str}\bigg\vert_{\vec{m}_{E_8}=0} = -\frac{(2\pi)^2}{24}[\epsilon_1\epsilon_2(|\nu|^2+|\nu|)-\epsilon_+^2|\nu|] \cdot Z_n^{E-str}\bigg\vert_{\vec{m}_{E_8}=0},
\end{equation}
where $ \epsilon_+ = \epsilon_1+\epsilon_2 $. This is most easily satisfied by requiring that each summand $ Z_\nu = D_\nu^{M9}D_{\nu\emptyset}^{M5} $ in Eq. \eqref{eq:Estringdomain} satisfy the same equation, so we conjecture that the following holds:
\begin{equation}
\partial_{E_2} Z_\nu^{E-str}\bigg\vert_{\vec{m}_{E_8}=0} = -\frac{(2\pi)^2}{24}[\epsilon_1\epsilon_2(|\nu|^2+|\nu|)-\epsilon_+^2|\nu|]\cdot Z_\nu^{E-str}\bigg\vert_{\vec{m}_{E_8}=0}.
\end{equation}
The Weyl[$ E_8 $]-invariant Jacobi forms $ A_{1,2,3,4,5}, \, B_{2,3,4,6} $ satisfy the following modular anomaly equation:
\begin{align}
\partial_{E_2} A_n(\tau;\vec{m}_{E_8}) &= -n\cdot\frac{(2\pi)^2}{24}\left(\sum_{i=1}^8 m^2_{E_8,i}\right)A_n(\tau;\vec{m}_{E_8}),\\
\partial_{E_2} B_n(\tau;\vec{m}_{E_8}) &= -n\cdot\frac{(2\pi)^2}{24}\left(\sum_{i=1}^8 m^2_{E_8,i}\right)B_n(\tau;\vec{m}_{E_8});
\end{align}
this leads us to guess the following form for the E-string holomorphic anomaly, for arbitrary values of $ \vec{m}_{E_8} $:
\begin{equation}
\partial_{E_2} Z_\nu^{E-str} = -\frac{(2\pi)^2}{24}\left[\epsilon_1\epsilon_2(|\nu|^2+|\nu|)-\epsilon_+^2|\nu|+|\nu|\left(\sum_{i=1}^8 m^2_{E_8,i}\right)\right]\cdot Z_\nu^{E-str}.\label{eq:EstrAnomaly}
\end{equation}

\item Finally, the elliptic genus for $ n $ E-strings is expected to be symmetric under exchange of $ \epsilon_1 $ and $ \epsilon_2 $. This is guaranteed by the fact that $ Z_\nu(\epsilon_1,\epsilon_2) $ and  $ Z_\nu (\epsilon_2,\epsilon_1) = Z_{\nu^t}(\epsilon_1,\epsilon_2) $ both appear in the expression for $ Z_{|\nu|}(\epsilon_1,\epsilon_2) $.
\end{itemize}

\noindent We also make the assumption that from the $ E_8 $ degrees of freedom (which for a single E-string are eight bosons compactified on the $ E_8 $ lattice) one obtains a factor of $ \eta^{8n} $ in the denominator of the elliptic genus of $ n $ E-strings. From this, and from the first three properties listed above, we can immediately write down the following ansatz for the left M9 domain wall:
\begin{equation}
D^{M_9, L}_{\nu} = \frac{N^L_{\nu}(\tau; \vec{m}_{E_8,L},\epsilon_1,\epsilon_2)}{\eta(\tau)^{8|\nu|} B^L_\nu(\tau;\epsilon_1,\epsilon_2)F^R_\nu(\tau;\epsilon_1,\epsilon_2,m)},\label{eq:M9ansatz}
\end{equation}
where\footnote{Up to a prefactor $ t^{-\frac{|| \nu^t||^2}{2}} $ which is needed to ensure that after gluing the factors of $ \xi_{\pm} $ combine correctly into theta functions.}
\begin{equation}
B^L_\nu(\tau;\epsilon_1,\epsilon_2) = \prod_{(i,j)\in\nu}\xi_+(\epsilon_1(\nu_i-j+1)-\epsilon_2(\nu^t_j-i))\xi_-(\epsilon_1(\nu_i-j)-\epsilon_2(\nu^t_j-i+1))
\end{equation}
and
\begin{equation}
F^R_\nu(\tau;\epsilon_1,\epsilon_2) = \prod_{(i,j)\in\nu}\theta_1(-m-\epsilon_1(\nu_i-j+1/2)+\epsilon_2(-i+1/2))/\eta
\end{equation}
are obtained by requiring that they combine correctly with the bosonic (that is, denominator) and fermionic (numerator) pieces of the M5 domain wall $ D^{M5}_{\nu\emptyset} $ (Equation \eqref{eq:DM5}).\newline

\noindent Likewise, we take the right M9 domain wall to be given by
\begin{equation} D^{M_9, R}_{\nu} = \frac{N^R_{\nu}(\tau;\vec{m}_{E_8,R},\epsilon_1,\epsilon_2)}{\eta(\tau)^{8|\nu|} B^R_\nu(\tau;\epsilon_1,\epsilon_2)F^L_\nu(\tau;\epsilon_1,\epsilon_2,m)},\label{eq:M9Ransatz}\end{equation}
where\footnote{Up to a prefactor $ q^{-\frac{|| \nu||^2}{2}} $.}
\begin{equation}
B^R_\nu(\tau;\epsilon_1,\epsilon_2) = \prod_{(i,j)\in\nu}\xi_-(\tau;\epsilon_1(\nu_i-j+1)-\epsilon_2(\nu^t_j-i))\xi_+(\tau;\epsilon_1(\nu_i-j)-\epsilon_2(\nu^t_j-i+1))
\end{equation}
and
\begin{equation}
F^L_\nu(\tau;\epsilon_1,\epsilon_2) = \prod_{(i,j)\in\nu}\theta_1(\tau;-m+\epsilon_1(\nu_i-j+1/2)-\epsilon_2(-i+1/2))/\eta(\tau).
\end{equation}

\noindent The transformation that exchanges left and right M5 domain walls leaves $ \epsilon_1 $ and $ \epsilon_2 $ fixed (see Equation \eqref{eq:leftright}). Therefore it is natural to expect that
\begin{equation}
N^R_{\nu}(\vec{m}_{E_8,R},\epsilon_1,\epsilon_2) = N^L_{\nu}(\vec{m}_{E_8,R},\epsilon_1,\epsilon_2) = N(\vec{m}_{E_8,R},\epsilon_1,\epsilon_2),
\end{equation}
so the only difference between the numerators of the left and right domain walls is that they depend on the fugacities for the corresponding $ E_8 $ group. The non-trivial task is to compute the numerator factor $ N^L_{\nu} $; later in this section we will use the remaining properties listed above to uniquely determine it in the case of one and two E-strings.\newline

\noindent Once the M9 domain walls have been computed, it should also be possible to express the heterotic string partition function in terms of them (now replacing every occurrence of mass parameter $ m $ with $ (\epsilon_4-\epsilon_3)/2 $, as is more appropriate in this context):
\begin{equation}
Z^n_{het} \sim \sum D_n^{M9,L}(\vec{m}_{E_8,L}) \cdot D_n^{M9,R}(\vec{m}_{E_8,R}).
\end{equation}
We will show how this works explicitly for one and two heterotic strings in the following sections. For now, let us list the properties that the elliptic genus for $ n $ heterotic strings, reviewed in Section \ref{sec:4}, is known to satisfy:

\begin{itemize}

\item The elliptic genus for $ n $ heterotic strings transforms with modular weight 0 under the  $ SL(2,\mathbb{Z}) $ modular transformation
\begin{equation}
(t,\epsilon_1,\epsilon_2,\epsilon_3,\epsilon_4,\tau) \to \left(t,\frac{\epsilon_1}{c\tau+d},\frac{\epsilon_2}{c\tau+d},\frac{\epsilon_3}{c\tau+d},\frac{\epsilon_4}{c\tau+d},\frac{a \tau + b}{c\tau + d}\right).
\end{equation}

\item The elliptic genus for $ n $ heterotic strings is invariant under pairwise exchange of $ \epsilon_i,\epsilon_j $, for any $ i,j=1\dots,4 $. 

\item The modular anomaly equation for the heterotic string is given by Eq. \eqref{eq:hetanomaly}:
\begin{equation}
	\frac{\partial Z^{\textrm{Het}}_n}{\partial E_2} = n\cdot\frac{(2\pi)^2}{24}\left(\sum_{i=1}^4 \epsilon_i^2 - \sum_{i=1}^{8} ((m^L_{E_8,i})^2+(m^R_{E_8,i})^2) \right) Z^{\textrm{Het}}_n~.
\end{equation}
\end{itemize}

\subsection{E-string holomorphic anomaly}
\label{sec:6b}

The purpose of this section is to demonstrate that the modular anomaly equation \eqref{eq:EstrAnomaly} for the E-strings
can be easily derived from the modular anomaly equations for heterotic and M-strings, by using our ansatz for the M9 domain walls, Equation \eqref{eq:M9ansatz}.  To see this, note that each summand appearing in the elliptic genus of $ n $ M-strings, Equation \eqref{eq:ZMstring}, has the form
\begin{equation}
\frac{F_{\nu}^L \cdot F_{\nu}^R}{B_\nu^L \cdot B_\nu^R},\label{eq:Mblock}
\end{equation}
where the explicit expressions for these factors is unimportant for the present discussion. Similarly, from our domain wall ansatz we expect each summand in the expression for $ Z^{E-str}_n $ to have the form
\begin{equation}
\frac{N_\nu^L(\tau;\vec{m}_{E_8,L},\epsilon_1,\epsilon_2)}{\eta^{8 n} (B_\nu^L \cdot B_\nu^R)},\label{eq:Eblock}
\end{equation}
and each summand in $ Z^{het}_n $ to have the form
\begin{equation} \frac{N(\vec{m}_{E_8,L},\epsilon_1,\epsilon_2)\cdot N(\vec{m}_{E_8,R},\epsilon_1,\epsilon_2)}{\eta^{16 n} (B^L \cdot B^R)(F^L\cdot F^R)}.\label{eq:Hblock}
\end{equation}
Note that, if we set $ \vec{m}_{E_8,R}=\vec{m}_{E_8,L} $, Eqn. \eqref{eq:Eblock} is the square root of the product between Equations \eqref{eq:Mblock} and \eqref{eq:Hblock}. Therefore, we expect that
\begin{equation}
\frac{1}{(2\pi)^2}\frac{1}{Z^{\textrm{E-str}}_n}\frac{\partial Z^{\textrm{E-str}}_n}{\partial E_2} = \frac{1}{(2\pi)^2}\frac{1}{2Z^{\textrm{M-str}}_n}\frac{\partial Z^{\textrm{M-str}}_n}{\partial E_2}+\frac{1}{(2\pi)^2}\left[\frac{1}{2Z^{\textrm{het}}_n}\frac{\partial Z^{\textrm{het}}_n}{\partial E_2}\right]\bigg\vert_{\vec{m}_{E_8,R}=\vec{m}_{E_8,L}}.
\end{equation}
Indeed, a short calculation reveals that the right hand side is given by
\begin{eqnarray}
\text{r.h.s. }&=&-\frac{n}{24}\left[\epsilon_1 \epsilon_2 n + \left(\left(\frac{\epsilon_3-\epsilon_4}{2}\right)^2 - \left(\frac{\epsilon_1+\epsilon_2}{2}\right)^2\right)\right]+\frac{n}{48}(\sum_{i=1}^4 \epsilon_i^2- 2\sum_{i=1}^{8} (m^L_{E_8,i})^2)\nonumber\\
&=&-\frac{n}{24}\left[\epsilon_1 \epsilon_2 n + \left(\left(\frac{\epsilon_3-\epsilon_4}{2}\right)^2 - \left(\frac{\epsilon_3+\epsilon_4}{2}\right)^2\right)\right]\nonumber\\
&&+\frac{n}{48}((\epsilon_1+\epsilon_2)^2+(\epsilon_3+\epsilon_4)^2-2(\epsilon_1\epsilon_2+\epsilon_3\epsilon_4)- 2\sum_{i=1}^{8} (m^L_{E_8,i})^2)\nonumber\\
&=&-\frac{n}{24}\left[\epsilon_1 \epsilon_2 n -\epsilon_3\epsilon_4\right]+\frac{n}{48}(2(\epsilon_1+\epsilon_2)^2-2(\epsilon_1\epsilon_2+\epsilon_3\epsilon_4)- 2\sum_{i=1}^{8} (m^L_{E_8,i})^2)\nonumber\\
&=&\frac{n}{24}\left[(\epsilon_1+\epsilon_2)^2-(n+1)\epsilon_1\epsilon_2-\left(\sum_{i=1}^8 (m^L_{E_8,i})^2\right)\right],
\end{eqnarray}
which is identical to the conjectural E-string modular anomaly of Eq. \eqref{eq:EstrAnomaly} that was obtained using completely different techniques!

\subsection{One E-string and one heterotic string}
\label{sec:6c}

We now turn to the explicit computation of the M9 domain wall for the partition $ \nu = \ydiagram{1} $, and show that the elliptic genera for a single E-string and the one for a single heterotic string can both be written in terms of it. The elliptic genus for a single E-string is known exactly: it is simply given by the torus partition function for eight bosons compactified on an internal $ E_8 $ lattice and four spacetime bosons:
\begin{equation}
Z_1^{\textrm{E-str}} = -\left(\frac{A_1(\vec{m}_{E_8,L})}{\eta^8}\right)\frac{\eta^{2}}{\theta_1(\epsilon_1)\theta_1(\epsilon_2)},
\end{equation}
where $ A_1(\vec{m}_{E_8,L}) = \Theta_{E_8}(\tau;\vec{m}_{E_8,L}) $ is the $ E_8 $ theta function.
If we make the ansatz
\begin{equation}
Z_1^{E-str} = D^{M9, L}_{\ydiagram{1}} D^{M5, R}_{\ydiagram{1}~\emptyset},
\end{equation}
where
\begin{equation}
D^{M5}_{\ydiagram{1}~\emptyset} = \frac{\theta_1(-m-\epsilon_+/2)\,\eta^{-1}}{\xi_-(\epsilon_1)\xi_+(-\epsilon_2)},
\end{equation}
we immediately see that
\begin{eqnarray}
D^{M9, L}_{\ydiagram{1}} &=& \left(\frac{A_1(\vec{m}_{E_8,L})}{\eta^8}\right)\frac{\eta}{\theta_1(-m-\epsilon_+/2)}\cdot\frac{1}{\xi_+(\epsilon_1)\xi_-(-\epsilon_2)}\nonumber\\
&=& \left(\frac{A_1(\vec{m}_{E_8,L})}{\eta^8}\right)\frac{\eta}{\theta_1(\epsilon_3)}\cdot\frac{1}{\xi_+(\epsilon_1)\xi_-(-\epsilon_2)}.
\end{eqnarray}
Recalling that under left-right exchange $ \epsilon_3 \leftrightarrow \epsilon_4 $, $ \vec{m}_{E_8,L}\leftrightarrow \vec{m}_{E_8,R}  $, and $\xi_{\pm}\leftrightarrow \xi_{\mp} $, we also find that 
\begin{equation}
D^{M9, R}_{\ydiagram{1}} = \left(\frac{A_1(\vec{m}_{E_8,R})}{\eta^8}\right)\frac{\eta}{\theta_1(\epsilon_4)}\cdot\frac{1}{\xi_-(\epsilon_1)\xi_+(-\epsilon_2)}.
\end{equation}
It is now straightforward to verify that combining the left and right M9 domain walls gives
\begin{equation}
Z^{\textrm{het}}_{1} = -\left(\frac{A_1(\vec{m}_{E_8,L})\times A_1(\vec{m}_{E_8,R})}{\eta^{16}}\right)\frac{\eta^4}{\theta_1(\epsilon_1)\theta_1(\epsilon_2)\theta_1(\epsilon_3)\theta_1(\epsilon_4)},
\end{equation}
which is precisely the elliptic genus for a single heterotic string, since
\begin{equation}
A_1(\vec{m}_{E_8,L})\times A_1(\vec{m}_{E_8,R}) = \Theta_{E_8\times E_8}(\tau;\vec{m}_{E_8,L},\vec{m}_{E_8,R}).
\end{equation}
\vspace{.0in}

\subsection{Two E-strings and two heterotic strings}
\label{sec:6d}

We now turn to the discussion of domain walls for two strings. Using these domain walls we will be able to deduce an exact expression for the elliptic genus of two E-strings; we will also be able to obtain an expression for the elliptic genus of two heterotic strings which is in agreement with the orbifolding formula. Before turning to computations, we would like to highlight a remarkable fact. As discussed in Section \ref{sec:4}, the heterotic strings do not form bound states, and therefore their elliptic genus can be computed by means of the Hecke transform; on the other hand, the E-strings, like the M-strings, do form bound states and therefore do not admit such a simple description. Nevertheless, we will see that the same building blocks -- the M9 and M5 domain walls -- can be used to compute the elliptic genera for two E-strings as well as for two heterotic strings.\\

\noindent Following the approach outlined at the beginning of the section, we start by making the following ansatz for the two E-string elliptic genus:
\begin{equation}
Z_2^{\textrm{E-str}} =  D_{\substack{\ydiagram{1,1}}}^{M9,L} D^{M5}_{\ydiagram{2}~\emptyset} +D_{\ydiagram{2}}^{M9,L}   D^{M5}_{\ydiagram{1,1}~\emptyset},
\end{equation}
where
\begin{equation}
D^{M5}_{\ydiagram{2}~\emptyset}= \frac{\theta_1(m+\epsilon_+/2)\eta^{-1}}{\xi_-(\epsilon_1)\xi_+(-\epsilon_2)}\frac{\theta_1(m+\epsilon_+/2+\epsilon_1)\eta^{-1}}{\xi_-(2\epsilon_1)\xi_+(\epsilon_1-\epsilon_2)}
\end{equation}
and
\begin{equation}
D^{M5}_{\ydiagram{1,1}~\emptyset}=\frac{\theta_1(m-\epsilon_+/2)\eta^{-1}}{\xi_+(\epsilon_1)\xi_-(-\epsilon_2)}\frac{\theta_1(m-\epsilon_+/2-\epsilon_2)\eta^{-1}}{\xi_+(-2\epsilon_2)\xi_-(\epsilon_1-\epsilon_2)}.
\end{equation}
This leads to the following ansatz for the M9 domain walls:
\begin{align}
D_{\ydiagram{2}}^{M9,L} &= \frac{N_{\ydiagram{2}}(\vec{m}_{E_8,L},\epsilon_1,\epsilon_2)/\eta^{16}}{\xi_+(\epsilon_1)\xi_-(-\epsilon_2)\xi_-(\epsilon_1-\epsilon_2)\xi_+(2\epsilon_1)\theta_1(m+\epsilon_+/2)\theta_1(m+\epsilon_+/2+\epsilon_1)\eta^{-2}},\\
D_{\ydiagram{1,1}}^{M9,L} &=  \frac{N_{\ydiagram{1,1}}(\vec{m}_{E_8, L},\epsilon_1,\epsilon_2)/\eta^{16}}{\xi_-(\epsilon_1)\xi_+(-\epsilon_2)\xi_+(\epsilon_1-\epsilon_2)\xi_-(-2\epsilon_2)\theta_1(m-\epsilon_+/2)\theta_1(m-\epsilon_+/2-\epsilon_2)\eta^{-2}}.
\end{align}
We expect that the two E-string elliptic genus can be written as
\begin{equation}
Z_2^{\textrm{E-str}} = -\frac{N_{\ydiagram{2}}(\vec{m}_{E_8,L},\epsilon_1,\epsilon_2)/\eta^{16}}{\theta_1(\epsilon_1)\theta_1(\epsilon_2)\theta_1(\epsilon_1-\epsilon_2)\theta_1(2\epsilon_1)\eta^{-4}}-\frac{N_{\ydiagram{1,1}}(\vec{m}_{E_8,L},\epsilon_1,\epsilon_2)/\eta^{16}}{\theta_1(\epsilon_1)\theta_1(\epsilon_2)\theta_1(\epsilon_2-\epsilon_1)\theta_1(2\epsilon_2)\eta^{-4}}.\label{eq:Z2Ansatz}
\end{equation}
To fix the numerator terms we exploit the following facts:
\begin{itemize}
\item Modular invariance of $ Z_{2}^{E-str} $ requires the modular weight of $ N_{\ydiagram{2}} $ and $ N_{\ydiagram{1,1}}$ to be 8 in order to cancel with the modular weight of the denominator (since $ \eta $ and $ \theta_1(z) $ have modular weight 1/2).
\item The numerator terms can be written as linear combinations of the three level-two Weyl[$ E_8 $]-invariant modular forms $ A_1^2, A_2,$ and $ B_2 $.
\item From the modular anomaly equation for $ Z_2^{E-str} $ one obtains
\begin{equation}
\frac{1}{N_{\ydiagram{2}}}\partial_{E_2} N_{\ydiagram{2}} = -\frac{(2\pi)^2}{12}\left[\epsilon_1^2+\left(\sum_{i=1}^8(m_{E_8,i}^L)^2\right)\right];
\end{equation}
the $ m_{E_8, L} $ terms in this equation are consistent with the fact that $ N_{\ydiagram{2}} $ can be expressed in terms of level 2 characters of affine $ E_8 $, while the $ -\epsilon_1^2/12 $ term indicates that $ N_{\ydiagram{2}} $ is a function of $ \epsilon_1$ and not $ \epsilon_2 $, and furthermore that it transforms with index 2 with respect to $ \epsilon_1 $ under modular transformations.
Since  $N_{\ydiagram{2}}(\tau;\vec{m}_{E_8,L},\epsilon_2,\epsilon_1) =  N_{\ydiagram{1,1}}(\tau;\vec{m}_{E_8,L},\epsilon_1,\epsilon_2)$, an analogous conclusion holds for $ N_{\ydiagram{1,1}} $, which is written in terms of level 2 $E_8$ characters and index two Jacobi forms with elliptic parameter $ \epsilon_2 $.
\end{itemize}

\noindent Concretely, this forces the numerator terms to have the following form:
\begin{align}
N_{\ydiagram{2}}(\tau;\vec{m}_{E_8,L},\epsilon_1,\epsilon_2) &=& A_1(\vec{m}_{E_8,L})^2 f_1(\tau;\epsilon_1) +  B_2(\vec{m}_{E_8,L}) f_2(\tau;\epsilon_1)+ A_2(\vec{m}_{E_8,L}) f_3(\tau;\epsilon_1), \label{eq:N2}\\
N_{\ydiagram{1,1}}(\tau;\vec{m}_{E_8,L},\epsilon_1,\epsilon_2) &=& A_1(\vec{m}_{E_8,L})^2 f_1(\tau;\epsilon_2) +  B_2(\vec{m}_{E_8,L}) f_2(\tau;\epsilon_2)+ A_2(\vec{m}_{E_8,L}) f_3(\tau;\epsilon_2),\label{eq:N11}
\end{align}
where $ f_1(\tau;\epsilon), f_2(\tau;\epsilon), f_3(\tau;\epsilon)  $ are Jacobi forms of index 2 with elliptic parameter $ \epsilon $, respectively of modular weight $0$, $2$ and $4$.\newline

\noindent{We now resort to the following fact about Jacobi forms \cite{Dabholkar:2012nd}}:\\\newline
{ \noindent\emph{The weak Jacobi forms with modular parameter $ \tau $ and elliptic parameter $ \epsilon $ of index $ k $ and even weight $ w $ form a polynomial ring which is generated by the four modular forms $ E_4(\tau), E_6(\tau), \phi_{0,1}(\epsilon,\tau), $ and $\phi_{-2,1}(\epsilon,\tau) $, where}
\[ \phi_{-2,1}(\epsilon,\tau) = -\frac{\theta_1(\epsilon;\tau)^2}{\eta^6(\tau)}\qquad \text{\emph{and}}\qquad
\phi_{0,1}(\epsilon,\tau) = 4\left[\frac{\theta_2(\epsilon;\tau)^2}{\theta_2(0;\tau)^2}+\frac{\theta_3(\epsilon;\tau)^2}{\theta_3(0;\tau)^2}+\frac{\theta_4(\epsilon;\tau)^2}{\theta_4(0;\tau)^2}\right]\]
\emph{are Jacobi forms of index 1, respectively of weight $ -2 $ and 0.}}\newline

\noindent Thus modularity implies that $ f_1, f_2, f_3 $ can be written as follows:
\begin{align} f_1(\epsilon) &= c_{1,1} \phi_{0,1}(\epsilon)^2+c_{1,2} E_4 \phi_{-2,1}(\epsilon)^2;\\
f_2(\epsilon) &= c_{2,1} E_4 \phi_{0,1}(\epsilon)\phi_{-2,1}(\epsilon)+c_{2,2} E_6 \phi_{-2,1}(\epsilon)^2;\\
f_3(\epsilon) &= c_{3,1} E_4\phi_{0,1}(\epsilon)^2+ c_{3,2} E_6 \phi_{0,1}(\epsilon)\phi_{-2,1}(\epsilon) +c_{3,3} E_4^2(\tau) \phi_{-2,1}(\epsilon)^2.
\end{align}
We now can determine the numerical coefficients $ c_{i,j} $ as follows: we use the results for the topological string free energy for up to $ n = 2 $ computed in \cite{Huang:2013yta} to calculate $ Z_2^{\textrm{E-str}}$ as an expansion in $ \epsilon_1\cdot\epsilon_2 $ and $ \epsilon_1+\epsilon_2 $, and match it against our ansatz \eqref{eq:Z2Ansatz}. We find that the terms in the free energy up to $ g + \ell = 2 $ are sufficient to uniquely fix all the coefficients in our expression.\footnote{In fact, the form of the numerator is constrained even further if we observe that, when we set $ \vec{m}_{E_8}\to 0 $, it should be given by a genuine holomorphic Jacobi form in $ \epsilon_1 $ or $ \epsilon_2 $ (not just a weakly holomorphic Jacobi form), as one expects from a unitary theory.} We find the following result:
\begin{align}
&N_{\ydiagram{2}}(\vec{m}_{E_8,L},\epsilon_1)= \frac{1}{576}\bigg[4 A_1^2(\phi_{0,1}(\epsilon_1)^2- E_4\phi_{-2,1}(\epsilon_1)^2)\hspace{2in}\nonumber\\
&\hspace{.1in}+3A_2(E_4^2\phi_{-2,1}(\epsilon_1)^2-E_6\phi_{-2,1}(\epsilon_1)\phi_{0,1}(\epsilon_1))+5 B_2(E_6\phi_{-2,1}(\epsilon_1)^2-E_4 \phi_{-2,1}(\epsilon_1)\phi_{0,1}(\epsilon_1))\bigg],
\end{align}
and
\begin{equation}
N_{\ydiagram{1,1}}(\vec{m}_{E_8,L},\epsilon_2) = N_{\ydiagram{2}}(\vec{m}_{E_8,L},\epsilon_2).
\end{equation}
In fact, in \cite{Huang:2013yta} the free energy was computed up to $ g + \ell = 3 $; we have checked that our domain wall expressions also match exactly with those coefficients; this provides a nontrivial check that we have found a formula for the two E-string elliptic genus which is exact to all orders in $ g $ and $ \ell $. Given the explicit formula for the two E-string elliptic genus, Equation \eqref{eq:Z2Ansatz}, one can easily check that the answer is not what one would have obtained using the Hecke transform:
\begin{align}
Z_{2}^{E-str}(\tau;\vec{m}_{E8}, \epsilon_1,\epsilon_2) &\neq \frac{1}{2}\left[Z_1^{E-str}(\tau;\vec{m}_{E8}, \epsilon_1,\epsilon_2)^2+Z_1^{E-str}(2\tau;2\vec{m}_{E8}, 2\epsilon_1,2\epsilon_2)\right. \nonumber\\
&+ \left. Z_1^{E-str}(\tau/2;\vec{m}_{E8}, \epsilon_1,\epsilon_2)+Z_1^{E-str}(\tau/2+1/2;\vec{m}_{E8}, \epsilon_1,\epsilon_2) \right].
\end{align}
This was to be expected, since the right hand side is not supposed to produce the right answer in contexts where the strings can form bound states.\\

\noindent We now would like to demonstrate that the M9 domain walls can also be used to compute the two heterotic string partition function. Note that
\begin{eqnarray}
~ & ~ & D_{\ydiagram{2}}^{M9,L}(\vec{m}_{E_8,L})D_{\ydiagram{2}}^{M9,R}(\vec{m}_{E_8,R}) \nonumber \\
~ & = & -\frac{N_{\ydiagram{2}} (\vec{m}_{E_8,L},\epsilon_1)N_{\ydiagram{2}} (\vec{m}_{E_8,R},\epsilon_1)}{\eta(\tau)^{24}\theta_1(\epsilon_1)\theta_1(\epsilon_2)\theta_1(\epsilon_3)\theta_1(\epsilon_4)\theta_1(2\epsilon_1)\theta_1(\epsilon_1-\epsilon_2)\theta_1(\epsilon_1-\epsilon_3)\theta_1(\epsilon_1-\epsilon_4)}\nonumber\\
~ & = & -\frac{N_{\ydiagram{2}} (\vec{m}_{E_8,L},\epsilon_1)N_{\ydiagram{2}} (\vec{m}_{E_8,R},\epsilon_1)}{\eta(\tau)^{24}\theta_1(\epsilon_1)\theta_1(\epsilon_2)\theta_1(\epsilon_3)\theta_1(\epsilon_4)\theta_1(2\epsilon_1)\theta_1(\epsilon_1-\epsilon_2)\theta_1(\epsilon_1-\epsilon_3)\theta_1(\epsilon_1-\epsilon_4)}
\end{eqnarray}
and
\begin{eqnarray} 
~ & ~ &D_{\ydiagram{1,1}}^{M9,L}(\vec{m}_{E_8,L})D_{\ydiagram{1,1}}^{M9,R}(\vec{m}_{E_8,R}) \nonumber \\
~ & = &- \frac{N_{\ydiagram{1,1}} (\vec{m}_{E_8,L},\epsilon_2)N_{\ydiagram{1,1}}(\vec{m}_{E_8,R},\epsilon_2)}{\eta(\tau)^{24}\theta_1(\epsilon_1)\theta_1(\epsilon_2)\theta_1(\epsilon_3)\theta_1(\epsilon_4)\theta_1(2\epsilon_2)\theta_1(\epsilon_2-\epsilon_1)\theta_1(\epsilon_2-\epsilon_3)\theta_1(\epsilon_2-\epsilon_4)}\nonumber\\
~ & = &- \frac{N_{\ydiagram{1,1}} (\vec{m}_{E_8,L},\epsilon_2)N_{\ydiagram{1,1}} (\vec{m}_{E_8,R},\epsilon_2)}{\eta(\tau)^{24}\theta_1(\epsilon_1)\theta_1(\epsilon_2)\theta_1(\epsilon_3)\theta_1(\epsilon_4)\theta_1(2\epsilon_2)\theta_1(\epsilon_2-\epsilon_1)\theta_1(\epsilon_2-\epsilon_3)\theta_1(\epsilon_2-\epsilon_4)}.
\end{eqnarray}
Here we run into a puzzle: we would naively have guessed that
\begin{equation}
Z_2^{\textrm{het}} \overset{?}{=} \,\,D_{\ydiagram{2}}^{M9,L}(\vec{m}_{E_8,L})D_{\ydiagram{2}}^{M9,R}(\vec{m}_{E_8,R})+ D_{\ydiagram{1,1}}^{M9,L}(\vec{m}_{E_8,L})D_{\ydiagram{1,1}}^{M9,R}(\vec{m}_{E_8,R});\label{eq:2hetDW}
\end{equation}
however, notice that this expression is not invariant under arbitrary exchanges of the four parameters $ \epsilon_{1,\dots,4} $, as we would expect from the heterotic string! We find instead that the first summand is invariant under arbitrary permutation of $ \epsilon_{2,3,4} $ while the second is invariant under any permutation of $ \epsilon_{1,3,4} $; furthermore, the two terms are exchanged by $ \epsilon_1\leftrightarrow \epsilon_2 $. The most natural remedy for this is to symmetrize the right hand side of Equation \eqref{eq:2hetDW}. This leads to the following formula for the elliptic genus for two heterotic strings:
\begin{equation} \label{eq:Hetstrres}
Z_2^{\textrm{het}} = D_{\ydiagram{2}}^{M9,L}(\vec{m}_{E_8,L})D_{\ydiagram{2}}^{M9,R}(\vec{m}_{E_8,R})+(\epsilon_1\leftrightarrow\epsilon_2)+(\epsilon_1\leftrightarrow\epsilon_3)+(\epsilon_1\leftrightarrow\epsilon_4).
\end{equation}
We find by direct comparison that this expression exactly matches with the orbifold formula for the elliptic genus for two heterotic strings, despite their completely different appearance!\footnote{We have checked this result up to powers of $ Q_\tau^8 $, with a generic choice of $ E_8 \times E_8 $ Wilson lines.}\newline

\noindent In order to highlight the novel properties of formula (\ref{eq:Hetstrres}), let us recall here the result of the orbifold formula of Section \ref{sec:4} specialized to the case of two heterotic strings:
\begin{eqnarray}
	Z_2^{\textrm{het}}(\tau,\vec{\epsilon}, \vec{m}_{E_8\times E_8})& = & \frac{1}{2}\left[\left(Z_1^{\textrm{het}}(\tau,\vec{\epsilon},\vec{m}_{E_8\times E_8})\right)^2 + Z_1^{\textrm{het}}(2\tau,2\vec{\epsilon},2\vec{m}_{E_8\times E_8})\right. \nonumber \\
	~ & ~ & + \left. Z_1^{\textrm{het}}(\frac{\tau}{2},\vec{\epsilon},\vec{m}_{E_8\times E_8})+Z_1^{\textrm{het}}(\frac{\tau+1}{2},\vec{\epsilon},\vec{m}_{E_8\times E_8})\right]. \nonumber \\ \label{eq:symhet2}
\end{eqnarray}
One can clearly see from this expression that the left and right $E_8$ masses are entangled in a non-trivial way. By this we mean that it is not possible to perferm independent $SL(2,\mathbb{Z})$ transformations on the left and right degrees of freedom, since under an $ SL(2,\mathbb{Z}) $ transformation the last three terms in Equation \eqref{eq:symhet2} transform into each other in a nontrivial way. In contrast to this expression (\ref{eq:Hetstrres}) is manifestly $SL(2,\mathbb{Z})$ invariant and one can perform independent modular transformations on the left and right degrees of freedom.

\subsection{Discussion of results}
\label{sec:6e}

We have derived expressions for one and two E-strings (\ref{eq:Z2Ansatz}) as well as for one and two heterotic strings (\ref{eq:Z2Ansatz}) in terms of domain wall building blocks. These expressions lead for the first time to an exact expression for the elliptic genus for two E-strings and a new representation for the elliptic genus of two heterotic strings. Our results have a few unusual and interesting properties on which we want to comment. First of all, observe that the well-known orbifold representation for heterotic strings given by (\ref{eq:Hetpfres}) has a very simple pole structure in the parameters $\epsilon_i$. On the other hand, the individual terms in Equation \eqref{eq:Hetstrres} have a more complicated pole structure. The agreement with the orbifold formula is a consequence of a nontrivial cancelation between the poles appearing in these terms.\\

\noindent We also wish to remark that it should be possible to give a direct physical interpretation to M9 domain wall expressions. This was the case for the M5 brane domain wall formula, which in \cite{Haghighat:2013gba} was shown to be equal to the open topological string partition function for a certain toric Calabi-Yau threefold. It would be interesting to determine whether the M9 domain wall formulas can also be related to the computation of the open topological string partition function on some specific Calabi-Yau geometry. A hint in this direction comes from the fact that the M9 expressions we computed have an integral expansion in the parameters $Q_{\tau}$, $q$, $t$ and $Q_m$, which is consistent with a BPS degeneracy interpretation of the expansion coefficients.\\

\noindent It remains to comment on the validity of our ansatz for three or more strings. One supporting argument we provided is that our domain wall picture leads to the correct holomorphic anomaly equation for the E-string from those of the M-string and heterotic string.\footnote{See also \cite{Ohmori:2014pca} for a computation of the anomaly polynomial of E-strings by using contributions from the M5 brane and M9 plane.} Furthermore, we have checked that for three heterotic strings the leading term of the expected result given by the orbifold formula is reproduced correctly by our ansatz. If this proves to be the case for any number of strings it should be possible to compute arbitrary E-string elliptic genera by fixing the appropriate M9 domain wall expressions through the knowledge of the heterotic string result. We intend to pursue this line of research in future work.

\section*{Acknowledgements}
We would like to thank D. Gaiotto, A. Hanany, H.-C. Kim, G. S. Ng, T. Rudelius and especially S. Kim for valuable discussions. We would like to thank the SCGP for hospitality during the 11th Simons Workshop on Mathematics and Physics, where this work was initiated. B.H. would also like to thank the Perimeter Institute for Theoretical Physics for hospitality during the workshop ``Supersymmetric Quantum Field Theories in Five and Six Dimensions''. This research was supported in part by Perimeter Institute for Theoretical Physics. Research at Perimeter Institute is supported by the Government of Canada through Industry Canada and by the Province of Ontario through the Ministry of Economic Development \& Innovation. The work of B.H. is supported by the NSF grant DMS-1159412. The work of C.V. is supported in part by NSF grant PHY-1067976.

\appendix
\section{Modular and Jacobi forms}
\label{sec:appendix}

In this appendix we collect several definitions and results related to (quasi-)modular and Jacobi forms that we make extensive use of in the main body of the text.

\subsection*{Modular forms}
We begin by defining an important class of holomorphic functions of the modular parameter $ \tau $, the Eisenstein series, which have the following series expansion:
\begin{equation}
E_{2k}(\tau) = 1 - \frac{4k}{B_{2k}}\sum_{n=1}^\infty \sigma_{2k - 1}(n)q^n,
\end{equation}
where $ q = e^{2\pi i \tau} $, $ B_{2k} $ are the Bernoulli numbers, defined as $ \sum_{k=0}^\infty B_{k} \frac{x^k}{k!} = \frac{x}{e^x-1} $, and $ \sigma_k(n) = \sum_{d | n} d^k $. For $ k > 1 $, the Eisenstein series $ E_{2k} $ transforms as a holomorphic modular form of weight $ 2k $, in the sense that
\begin{equation}
E_{2k}\left(\frac{a \tau + b}{c\tau + d}\right) = (c\tau + d)^{2k} E_{2k}(\tau), \qquad \begin{pmatrix}a & b\\ c & d\end{pmatrix}\in SL(2,\mathbb{Z});
\end{equation}
For $ k = 1 $, the Eisenstein series $ E_2(\tau) $ is modular up to an anomalous term:
\begin{equation}
E_{2}\left(\frac{a \tau + b}{c\tau + d}\right) = (c\tau + d)^{2} E_{2}(\tau)-\frac{6ic}{\pi} (c\tau + d), \qquad \begin{pmatrix}a & b\\ c & d\end{pmatrix}\in SL(2,\mathbb{Z}).
\end{equation}

The anomalous term can be removed by defining an alternative form of the Eisenstein series,
\begin{equation}
\widehat{E}_2(\tau,\overline{\tau}) = E_2(\tau) - \frac{6i}{\pi (\tau - \overline{\tau})},
\end{equation}
at the cost of introducing a mild dependence on the anti-holomorphic parameter $ \overline{\tau} $. Unlike $ E_2(\tau) $, $ \widehat{E}_2(\tau,\overline\tau) $ transforms as an honest weight-two modular form:
\begin{equation}
\widehat{E}_{2}\left(\frac{a \tau + b}{c\tau + d},\frac{a \overline\tau + b}{c\overline\tau + d}\right) = (c\tau + d)^{2} \widehat{E}_{2}(\tau,\overline\tau), \qquad \begin{pmatrix}a & b\\ c & d\end{pmatrix}\in SL(2,\mathbb{Z}).
\end{equation}

\noindent The space of quasi-holomorphic $ SL(2,\mathbb{Z}) $ modular forms, that is, modular forms which are polynomials in $ \text{Im}(\tau)^{-1} $ with coefficients which are holomorphic functions of $ \tau $, is a polynomial ring which is generated by
\begin{equation}
\widehat E_2(\tau,\overline\tau),\, E_4(\tau),\text{ and } E_6(\tau);
\end{equation}
similarly,
\begin{equation}
E_2(\tau),\, E_4(\tau),\text{ and } E_6(\tau)
\end{equation}
generate the polynomial ring of $ SL(2,\mathbb{Z}) $ quasi-modular forms, defined simply by taking the holomorphic part of quasi-holomorphic modular forms.\\

\noindent Another function we will make extensive use of is the Dedekind eta function
\begin{equation}
\eta(\tau) = e^{\frac{\pi  i \tau}{12}}\prod_{n=1}^\infty (1-e^{2\pi i n \tau}),
\end{equation}
which, up to a phase, transforms as a weight-1/2 modular form:
\begin{equation}
\eta(\tau + 1) = e^{\frac{\pi i}{12}}\eta(\tau),\qquad \eta(-1/\tau) = e^{-\pi i/2} \tau^{1/2}\eta(\tau).
\end{equation}

\subsection*{Jacobi forms}
We now turn to a brief discussion of Jacobi forms, which are functions of a modular parameter $ \tau $ and an elliptic parameter $ z $. Under a modular transformation parametrized by $ \begin{pmatrix}a & b \\ c & d\end{pmatrix} \in SL(2,\mathbb{Z}) $, Jacobi forms transform as follow:
\begin{equation}
\phi\left(\frac{a \tau + b}{c\tau + d};\,\frac{z}{c\tau + d}\right) = (c\tau + d)^k e^{\frac{2\pi i m c z^2}{c\tau + d}}\phi(\tau;z);\end{equation}
also, under translations of $ z $ by $ \mathbb{Z} \tau + \mathbb{Z} $ they transform as follows:
\begin{equation}
\phi(\tau;z+\lambda \tau + \mu) = e^{-2\pi i m(\lambda^2 \tau + 2\lambda z)}\phi(\tau;z), \qquad \lambda, \mu \in \mathbb{Z}.\end{equation}
The two numbers $ k $ and $ m $ are referred to respectively as the weight and index of the modular form. Jacobi forms have a Fourier expansion of the form
\begin{equation}
\sum_{n, r} c(n,r) e^{2\pi i n\tau}e^{2\pi i r z}.
\end{equation}
One usually requires the coefficients $ c(n,r) $ of Jacobi forms to vanish for $ r^2 > 4 m n $; imposing the less strict condition that $ c(n,r) = 0 $ if $ n < 0 $ leads to a larger class of functions denoted as weak Jacobi forms.\\

\noindent A prominent example of Jacobi form is the Jacobi theta function
\begin{equation}
\theta_1(\tau;z) = -i e^{\pi i\tau/6}e^{\pi i z}\eta(\tau)\prod_{k=1}^\infty (1-e^{2\pi i k \tau}e^{2\pi i z})(1-e^{2\pi i (k-1) \tau}e^{-2\pi i iz}),
\end{equation}
which has weight $ 1/2 $ and index $ 1/2 $; it satisfies the property that
\[ \partial_{E_2} \theta_1(\tau;z) = \frac{(2\pi i z)^2}{24}\theta_1(\tau;z).\]
Closely related to $ \theta_1(z, \tau) $ are the functions
\begin{align}
\theta_2(\tau;z) &= \theta_1(\tau;z+1/2),\\
\theta_3(\tau;z) &= e^{\pi i z + \pi i \tau/4}\theta_1(\tau;z+1/2+\tau/2),\\
\theta_4(\tau;z) &= -ie^{\pi i z + \pi i \tau/4}\theta_1(\tau;z+\tau/2).
\end{align}

\noindent In the main text we make use of the following result concerning weak Jacobi forms of even weight (see for example \cite{Dabholkar:2012nd}):\\

{ \noindent\emph{The weak Jacobi forms with modular parameter $ \tau $ and elliptic parameter $ \epsilon $ of index $ k $ and even weight $ w $ form a polynomial ring which is generated by the four modular forms $ E_4(\tau), E_6(\tau), \phi_{0,1}(\tau;z), $ and $\phi_{-2,1}(\tau;z) $, where}
\[ \phi_{-2,1}(\tau;z) = -\frac{\theta_1(\tau;z)^2}{\eta^6(\tau)}\qquad \text{\emph{and}}\qquad
\phi_{0,1}(\tau;z) = 4\left[\frac{\theta_2(\tau;z)^2}{\theta_2(\tau;0)^2}+\frac{\theta_3(\tau;z)^2}{\theta_3(\tau;0)^2}+\frac{\theta_4(\tau;z)^2}{\theta_4(\tau;0)^2}\right]\]
\emph{are Jacobi forms of index 1, respectively of weight $ -2 $ and 0.}}\newline

\noindent It is important to note that
\[ \partial_{E_2}\phi_{-2,1}(\tau;z) = \frac{(2\pi i z)^2}{12}\phi_{-2,1}(\tau;z),\qquad \partial_{E_2}\phi_{0,1}(\tau;z) = \frac{(2\pi i z)^2}{12}\phi_{0,1}(\tau;z).\]

\noindent Lastly, we define the $ n $-th Hecke operator $T_n$ by its action on a weak Jacobi form $f(\tau,z)$ of weight $k$ as
\begin{equation}
	T_n f(\tau;z) = n^{k-1} \sum_{\stackrel{ad=n}{a,d>0}} \frac{1}{d^k} \sum_{b~ (\textrm{mod} ~d)} f\left(\frac{a\tau+b}{d};a z\right).
\end{equation}
Under this transformation, weak Jacobi forms of weight $ k $ and index $ m $ are mapped to weak Jacobi forms of weight $ k $ and index $ n\,m $.

\subsection*{Weyl invariant Jacobi forms for $ E_8 $}
We conclude this appendix by mentioning a class of multivariate Jacobi forms which are invariant under the Weyl group of $ E_8 $. These functions depend on the modular parameter $ \tau $ as well as eight parameters $ m_{E_8}^{1,\dots, 8} $, and are organized in two classes:
\begin{equation}
A_1,\, A_2,\, A_3,\, A_4,\, A_5 \qquad\text{ and }\qquad B_2,\, B_3,\, B_4,\, B_6.
\end{equation}
In the limit $ \vec{m}_{E_8} \to 0 $, these functions reduce to Eisenstein series:
\begin{equation}
A_i(\tau; \vec{m}_{E_8}) \to E_4(\tau); \qquad B_i(\tau;\vec{m}_{E_8}) \to E_6(\tau).
\end{equation}
Furthermore, one has:
\begin{align}
\partial_{E_2} A_n(\tau;\vec{m}_{E_8}) &= -n\cdot\frac{(2\pi)^2}{24}\left(\sum_{i=1}^8 m^2_{E_8,i}\right)A_n(\tau;\vec{m}_{E_8}),\\
\partial_{E_2} B_n(\tau;\vec{m}_{E_8}) &= -n\cdot\frac{(2\pi)^2}{24}\left(\sum_{i=1}^8 m^2_{E_8,i}\right)B_n(\tau;\vec{m}_{E_8}).
\end{align}
It is known that any Jacobi form which is given by a linear combination of characters of affine $ E_8 $ representations and is invariant under the Weyl group of $ E_8 $ can be written as a polynomial in $ A_{1,2,3,4,5},B_{2,3,5} $. The superscript of these functions indicates the amount by which they contribute to the level of the affine $E_8$ character. Thus, for example, any Weyl-invariant Jacobi form which is a combination of level $ 2 $ characters of affine $ E_8 $ can be written as a linear combination of $ A_1(\tau;\vec{m}_{E_8})^2, A_2(\tau;\vec{m}_{E_8}), $ and $ B_2(\tau;\vec{m}_{E_8}) $.\\

\noindent The simplest of these functions, $ A_1(\vec{m}_{E_8},\tau) $, is equal to the $ E_8 $ theta function
\begin{equation}
\Theta_{E_8}(\tau;\vec{m}_{E_8}) = \sum_{\vec{k}\in \Gamma_{E_8}}\exp(\pi i \tau \vec{k}\cdot \vec{k}+2\pi i \vec{m}_{E_8}\cdot\vec{k}) = \frac{1}{2}\sum_{k=1}^4\prod_{\ell=1}^8 \theta_{k}(\tau;m_{E_8}^\ell), 
\end{equation}
where $ \vec{k} $ runs over the points of the $E_8$ lattice $ \Gamma_{E_8} $. The other eight $ E_8 $ Jacobi functions can be defined starting from $ A_1(\vec{m}_{E_8},\tau) $, as discussed in more detail in \cite{Sakai:2011xg} and \cite{Huang:2013yta}.\\

\noindent Finally, it is worth mentioning the $ E_8\times E_8 $ theta function, which depends on sixteen parameters $ \vec{m}_{E_8\times E_8} $ and is defined as
\begin{equation}
\Theta_{E_8\times E_8}(\tau;\vec{m}_{E_8\times E_8}) = \sum_{\vec{k}\in \Gamma_{E_8\times E_8 }}\exp(\pi i \tau \vec{k}\cdot \vec{k}+2\pi i \vec{m}_{E_8\times E_8}).
\end{equation}
Since $ \Gamma_{E_8\times E_8} = \Gamma_{E_8} \oplus \Gamma_{E_8} $, one can pick a basis where
\begin{equation}
\vec{m}_{E_8\times E_8,1},\dots, \vec{m}_{E_8\times E_8,8} = \vec{m}_{E_8,L}
\end{equation}
only have nonzero product with the first $ \Gamma_{E_8} $ factor, while
\begin{equation}
\vec{m}_{E_8\times E_8,9},\dots, \vec{m}_{E_8\times E_8,16} = \vec{m}_{E_8,R}
\end{equation}
only have nonzero product with the second $ \Gamma_{E_8} $ factor. It is thus clear that
\begin{equation}
\Theta_{E_8\times E_8}(\tau;\vec{m}_{E_8\times E_8}) = \Theta_{E_8}(\tau;\vec{m}_{E_8,L})\Theta_{E_8}(\tau;\vec{m}_{E_8,R}) = A_1(\tau; \vec{m}_{E_8,L}) A_1(\tau; \vec{m}_{E_8,R}).
\end{equation}

\bibliography{references}

\end{document}